\documentclass[preprint, aps, prx, superscriptaddress]{revtex4}

\usepackage[T1]{fontenc} 
\usepackage{xr}
\usepackage{amsmath,amssymb,mathrsfs}
\usepackage{latexsym}
\usepackage{hyperref}
\usepackage{natbib}
\usepackage{graphicx}
\usepackage{amsfonts}
\usepackage[dvipsnames]{xcolor}
\usepackage{bm}
\usepackage{url}
\usepackage{soul}
\usepackage{cancel}
\usepackage{cleveref}
\usepackage{pgfplots,tikz}
\usepackage{xfrac}
\usepackage{pdfpages}
\usepackage{orcidlink}
\usepackage{lipsum}

\usepackage{color}
\usepackage{changes}
\pgfplotsset{compat=1.18}
\hypersetup{
    colorlinks = true,
    linkcolor = MidnightBlue,
    citecolor = MidnightBlue,
    linkbordercolor = {white},
    urlcolor = RoyalBlue
}

\crefname{equation}{Eq.}{Eqs.}
\Crefname{equation}{Equation}{Equations}
\crefname{figure}{Fig.}{Figs.}
\Crefname{figure}{Figure}{Figures}
\crefname{section}{Sec.}{Secs.}
\Crefname{section}{Section}{Sections}
\crefname{appendix}{Appendix}{Apps.}
\Crefname{appendix}{Appendix}{Apps.}
\crefname{paragraph}{Sec.}{Secs.}
\crefname{table}{Table}{Tables}

\newcommand{\textalert}[1]{}

\newcommand{\ket}[1]{\left|#1\right\rangle}
\newcommand{\bra}[1]{\left\langle#1\right|}

\newcommand{\braket}[2]{\bigl\langle#1\bigl|\bigr.#2\bigr\rangle}
\newcommand{\ketbra}[2]{\bigl|\bigr.#1\bigr\rangle\bigl\langle#2\bigr|}

\def\ie{\emph{i.e.},\ }

\definechangesauthor[name=Ivano, color=Magenta]{ita}
\definechangesauthor[name=Alberto, color=Orange]{ab}
\definechangesauthor[name=Anthony, color=RoyalBlue]{ag}
\definechangesauthor[name=Werner, color=ForestGreen]{wd}

\def\upuppath{\tikz[scale=0.25, baseline=.1ex,every node/.style={circle,fill=black!40, scale=0.5}]{
  \node (n0) at (0,0){};
  \node (n1) at (1,1){};
  \node (n2) at (2,2){};
  \foreach \from/\to/\name [count=\i] in {n0/n1,n1/n2}{
    \draw (\from) -- (\to) [color=gray!40, ultra thick] node {};
    }
}}

\def\downdownpath{\tikz[scale=0.25, baseline=.1ex,every node/.style={circle,fill=black!40, scale=0.5}]{
  \node (n0) at (0,2){};
  \node (n1) at (1,1){};
  \node (n2) at (2,0){};
  \foreach \from/\to/\name [count=\i] in {n0/n1,n1/n2}{
    \draw (\from) -- (\to) [color=gray!40, ultra thick] node {};
    }
}}

\def\updownpath{\tikz[scale=0.25, baseline=.1ex,every node/.style={circle,fill=black!40, scale=0.5}]{
  \node (n0) at (0,0){};
  \node (n1) at (1,1){};
  \node (n2) at (2,0){};
  \foreach \from/\to/\name [count=\i] in {n0/n1,n1/n2}{
    \draw (\from) -- (\to) [color=gray!40, ultra thick] node {};
    }
}}

\def\downuppath{\tikz[scale=0.25, baseline=.1ex,every node/.style={circle,fill=black!40, scale=0.5}]{
  \node (n0) at (0,1){};
  \node (n1) at (1,0){};
  \node (n2) at (2,1){};
  \foreach \from/\to/\name [count=\i] in {n0/n1,n1/n2}{
    \draw (\from) -- (\to) [color=gray!40, ultra thick] node {};
    }
}}

\begin{document}


\title{Quantum computing in spin-adapted representations for efficient simulations of spin systems}

\author{Anthony Gandon\,\orcidlink{0009-0004-3948-7239}}
\email{anthony.gandon@ibm.com}
\affiliation{IBM Quantum, IBM Research - Zurich, Säumerstrasse 4, 8803 Rüschlikon, Switzerland}
\affiliation{Institute for Theoretical Physics, ETH Zürich, Wolfgang-Pauli-Str. 27, 8093 Zürich, Switzerland}

\author{Alberto Baiardi\,\orcidlink{0000-0001-9112-8664}}
\affiliation{IBM Quantum, IBM Research - Zurich, Säumerstrasse 4, 8803 Rüschlikon, Switzerland}

\author{Max Rossmannek\,\orcidlink{0000-0003-1725-9345}}
\affiliation{IBM Quantum, IBM Research - Zurich, Säumerstrasse 4, 8803 Rüschlikon, Switzerland}

\author{Werner Dobrautz\,\orcidlink{0000-0001-6479-1874}}
\thanks{Current address: Center for Advanced Systems Understanding, Helmholtz-Zentrum Dresden-Rossendorf e.V. and Center for Scalable Data Analytics and Artificial Intelligence Dresden/Leipzig, Germany}
\affiliation{Department of Chemistry and Chemical Engineering, Chalmers University of Technology, 41296 Gothenburg, Sweden}


\author{Ivano Tavernelli\,\orcidlink{0000-0001-5690-1981}}
\affiliation{IBM Quantum, IBM Research - Zurich, Säumerstrasse 4, 8803 Rüschlikon, Switzerland}

\begin{abstract}
Exploiting inherent symmetries is a common and effective approach to speed up the simulation of quantum systems.
However, efficiently accounting for non-Abelian symmetries, such as the $SU(2)$ total-spin symmetry, remains a major challenge.
In fact, expressing total-spin eigenstates in terms of the computational basis can require an exponentially large number of coefficients.
In this work, we introduce a novel formalism for designing quantum algorithms directly in an eigenbasis of the total-spin operator.
Our strategy relies on the symmetric group approach in conjunction with a truncation scheme for the internal degrees of freedom of total-spin eigenstates.
For the case of the antiferromagnetic Heisenberg model, we show that this formalism yields a hierarchy of spin-adapted Hamiltonians, for each truncation threshold, whose ground-state energy and wave function quickly converge to their exact counterparts, calculated on the full model.
These truncated Hamiltonians can be encoded with sparse and local qubit Hamiltonians that are suitable for quantum simulations.
We demonstrate this by developing a state-preparation schedule to construct shallow quantum-circuit approximations, expressed in a total-spin eigenbasis, for the ground states of the Heisenberg Hamiltonian in different symmetry sectors.
\end{abstract}

\maketitle

\newpage

\section{Introduction}
\label[section]{introduction}

The Heisenberg model~\cite{Heisenberg1928ZurTheorieFerromagnetismus, Dirac1997QuantumMechanicsManyelectron, Dirac1997TheoryQuantumMechanics} has been widely used in solid-state physics to study the magnetic properties of physical systems. 
The corresponding Hamiltonian yields a simplified quantum-mechanical description of localized spins interacting on a discrete lattice. 
The model captures the phenomenology of many strongly correlated systems, such as transition metal materials~\citep{Tazhigulov2022SimulatingModelsChallenginga, Chen2022_Heisenberg-MnTorus}.
Despite its simplicity, exact solutions of the Heisenberg Hamiltonian in the thermodynamic limit are limited to exceptional geometries, with the one-dimensional chain of spin-$1/2$ particles being a prime example of an analytically solvable model~\citep{Bethe1931ZurTheorieMetalle,Karabach1997IntroductionBetheAnsatza}.
For finite-size lattices, exact numerical simulations are limited to systems of up to approximately 50 spins due to the exponential scaling of the Hilbert space size with the number of lattice sites~\cite{Lscher2009, Nataf2014, Luchli2019}.
However, efficient, highly accurate heuristic simulation techniques that either make use of the locality of the Hamiltonian or leverage its symmetries can target even larger systems.
The ground-state of one-dimensional (1D) and quasi-1D ladder-like lattices are characterized by an area-law entanglement~\cite{Schollwoeck2005_Review, Schollwoeck2011_Review, Baiardi2020_Review, Verstraete2006, Eisert2010}, and can therefore be accurately approximated with matrix product state-based methods, such as the density matrix renormalization group approach.
Methods that are able to target more complex topologies, such as neural-network-~\cite{Carleo2017, Choo2020, Melko2019, Nomura2017} and multi-dimensional tensor-networks~\cite{Vidal2007, Murg2009, Ors2014}, are continuously being developed. 
On frustration-free, bipartite lattices, the problem does not exhibit a sign problem~\cite{Lee1984, Loh1990, Manousakis1991, Troyer2005}, and therefore quantum Monte Carlo (QMC) approaches can provide accurate results~\cite{Lee1984, Hirsch1985, Ghanem2021, Evertz2003, Hirsch1986, Foulkes2001, Guther2020}.

In recent years, quantum computing has emerged as a promising candidate for estimating properties of spin Hamiltonians beyond the limit of exact diagonalization~\cite{Kim2023EvidenceUtilityQuantum, Robledo-Moreno2024ChemistryExactSolutions, Yoshioka2024DiagonalizationLargeManybodya, Chowdhury2024EnhancingQuantumUtilityb}.
It relies on mapping the quantum-mechanical spins to engineered quantum degrees of freedom.
As such, quantum computing provides a natural framework for studying dynamical properties of real-time evolved quantum states and static ground state properties of quantum systems. 

A further strategy to tame exponentially-scaling computational costs is to design algorithms that account for the system inherent symmetries. 
Various solutions for resolving non-Abelian symmetries, such as the spin-rotational symmetry of the Heisenberg model, have been developed in classical computation algorithms, including memory efficient spin-adapted exact diagonalization studies~\citep{flockeSymmetricgroupApproachStudies1997, flockeSymmetricGroupApproachHeisenberg2002, Heitmann2019, Ido2024, LiManni2023}, non-abelian DMRG~\citep{Tatsuaki2000, McCulloch2002NonAbelianDensityMatrixa, Sharma2012SpinadaptedDensityMatrix, Singh2012, Keller2016_SpinAdapted}, to recent advances on spin-adapted quantum Monte Carlo (QMC) variants~\citep{Vieijra2021ManybodyQuantumStates, dobrautzEfficientFormulationFull2019, Dobrautz2021, Yun_2021, dobrautzCombinedUnitarySymmetric2022}.
In the field of quantum computing, symmetry-adapted approaches have been limited so-far mainly to Abelian symmetries.
A formalism was developed for binary Abelian symmetries~\citep{Bravyi2017TaperingQubitsSimulateb}, and has been later extended to other types of Abelian symmetries~\citep{Cheng2023UnleashingQuantumSimulation, Picozzi2023SymmetryadaptedEncodingsQubit, Setia2020ReducingQubitRequirements, Shee2022QubitefficientEncodingScheme}. 
Within this formalism, the degrees of freedom in the symmetry subspace are mapped to the qubit register. This allows for a reduction of the size of the computational space and ensures that errors occurring during the experimental realization of any quantum algorithms on noisy hardware do not break the conservation of symmetry quantum numbers. 

For non-Abelian symmetries, expressing a wave function in a symmetry-adapted basis becomes more challenging.
For this reason, recent works have focused on developing algorithms that enforce the conservation of symmetry quantum numbers, but working in \textit{standard}, non-spin-adapted bases. 
For instance, shallow quantum circuits have been derived to initialize low-energy eigenstates of the total-spin operator~\citep{Bartschi2019DeterministicPreparationDickeb, Piroli2024ApproximatingManybodyQuantumb, Raveh2024DickeStatesMatrixa, Sugisaki2019OpenShellElectronic, Carbone2022QuantumCircuitsPreparation, sugisakiQuantumChemistryQuantum2016b, Marti-Dafcik2024SpinCouplingAll, Moerchen2024_ClassificationElectronicStructures}, which can be used as state-preparation primitives. 
As these eigenstates have been shown to have large overlaps with the true ground states, they can be used as input for the quantum phase estimation algorithm~\cite{Nielsen_Chuang2010}.
Spin symmetry has been also leveraged to construct quantum number-preserving ansätze~\citep{Gard2020EfficientSymmetrypreservingState, Anselmetti2021LocalExpressiveQuantumnumberpreserving, East2023AllYouNeed, Burton2024AccurateGateefficientQuantum, Berry2024_RapidInitialStatePreparation}, with applications in the context of variational optimization. 
Finally, strategies have been developed for projecting a wave function encoded as a quantum circuit on a given spin symmetry subspace using quantum measurements, either in classical post-processing or using ancillary qubits~\citep{Seki2020SymmetryadaptedVariationalQuantum, Bastidas2024UnificationFiniteSymmetries, Lacroix2020SymmetryAssistedPreparationEntangled, Tsuchimochi2020_SpinProjection}

In this work, we propose an alternative framework where quantum algorithms are directly formulated in a spin-adapted basis.
We use a truncation of the internal total-spin quantum numbers to define a hierarchy of increasingly more accurate encodings of the spin-adapted subspace on quantum registers with constant local dimension $d=2$, \ie qubit registers.
Although it is in principle approximate, such a truncation scheme has been shown to yield very accurate representation of the full Hamiltonian based on classical VMC calculations~\cite{Vieijra2021ManybodyQuantumStates}.
Notably, our approach avoids explicitly implementing the general unitary basis change between the computational $\hat{\mathbf{s}}_z$-eigenbasis and the $\hat{\mathbf{S}}^2$-eigenbasis using, e.g., the quantum Schur transformation~\citep{Bacon2006EfficientQuantumCircuits}.
This transformation is impractical to implement on most quantum hardware, as it relies on encoding spin-adapted basis states in a register of qudits, whose length and local dimensions grow with the system size. 
Within our framework, we show that the qubit encodings of the Hamiltonian in the truncated spin-adapted bases are sparse and local, which makes them suitable for designing quantum algorithms tailored to currently available hardware with limited connectivity. We apply our strategy to the one-dimensional antiferromagnetic Heisenberg Hamiltonian and demonstrate a resource-efficient adiabatic state-preparation strategy for preparing its ground states in various symmetry sectors and in a total-spin eigenbasis.

The paper is organized as follows: 
In \cref{sec:spin_adapted_bases}, we summarize how spin-adapted bases are constructed from \emph{spin-coupling schemes} for systems with global $SU(2)$ symmetry.
Moreover, we introduce a truncation scheme of the full spin-adapted basis, inspired by Ref.~\cite{Vieijra2020RestrictedBoltzmannMachinesa}, which will be a key component of our formalism.
For the nearest-neighbor antiferromagnetic one-dimensional Heisenberg Hamiltonian, we show that the truncated spin-adapted subspaces accurately encode the low-energy subspace of the full model.
In \cref{sec:matrix_representation} we present the symmetric group approach (SGA) for determining the representation of spin-free Hamiltonians on the truncated spin-adapted subspace. We derive sparse and local qubit encodings for each of these truncated spin-adapted Hamiltonians in \cref{sec:qudit_to_qubit_mapping}, and further show in \cref{sec:time_evolution_truncated_hamiltonians}, that the dynamical real-time properties of this hierachy of qubit Hamiltonians quickly converge to the exact dynamical properties of the full model.
Finally, we leverage this observation in \cref{sec:adiabatic_ground_state_preparation} to prepare ground state approximations of the Heisenberg Hamiltonian in a spin-adapted basis, using shallow quantum circuits and adiabatic state-preparation schedules.

\section{Heisenberg Hamiltonian and spin-adapted bases}
\label[section]{sec:spin_adapted_bases}

We study the Heisenberg Hamiltonian describing the exchange interaction between localized quantum-mechanical spins on a lattice
\begin{align}
    \hat{\mathbf{H}}&= \sum_{\langle i,j \rangle } J_{ij} \hat{\mathbf{s}}_i \cdot \hat{\mathbf{s}}_j \, ,
\end{align}
where $\hat{\mathbf{s}}_i$ are spin operators with local quantum numbers, $s_i=\{1/2, 1, \dots \}$. 
We consider a finite system of $N$ spin-$1/2$s interacting through an isotropic antiferromagnetic interaction between nearest-neighbor spins $\langle i,j \rangle$, corresponding to an interaction strength $J_{ij} = J >0$. 
We will focus on the calculation of the ground state of the Heisenberg Hamiltonian in one dimension.
Although this state can be represented through the Bethe ansatz~\citep{Bethe1931ZurTheorieMetalle}, we will follow a different formalism that enables integrating symmetries in quantum algorithms. 
The proposed approach could be used in more general applications, for example, where interactions between non-adjacent sites are included.

\subsection{Symmetries}
The Heisenberg Hamiltonian possesses a spin-rotational $SU(2)$-symmetry corresponding to the conservation of the total spin operator $\hat{\mathbf{S}}^2$, as well as a $U(1)$-symmetry associated with the conservation of the total magnetization operator $\hat{\mathbf{S}}_z$. 
These spin operators commute with the Hamiltonian, \ie 
\begin{equation}
    [\hat{\mathbf{H}}, \hat{\mathbf{S}}^2] = [\hat{\mathbf{H}}, \hat{\mathbf{S}}_z] = 0 \, .
\end{equation}
Consequently, a symmetry-adapted basis block diagonalizes the Hamiltonian matrix according to the conserved quantum numbers. This property can be leveraged to reduce the dimensionality of the exponentially-large Hilbert space of $N$ spins when solving the corresponding eigenvalue problem. For example, the total magnetization operator $\hat{\mathbf{S}}_z$ splits the set of $2^N$ $\hat{\mathbf{s}}_z$-basis states, defined from the eigenvalues of the local magnetization as $\{\ket{m_1, m_2,\cdots, m_N}| m_i\in\{0,1\}\}$, into total magnetization subspaces according to their global eigenvalue $M = \sum_{i=1}^{N} m_i$. 

The $\hat{\mathbf{S}}_z$ block diagonalization is especially simple as the total magnetization is diagonal in the $\hat{\mathbf{s}}_z$-basis, which is the standard computational basis. Generally, a basis rotation is necessary to express the Hamiltonian matrix in symmetry-adapted bases. This applies to the spin-rotational $SU(2)$ symmetry, as non-abelian symmetry eigenstates cannot be written as tensor products of eigenstates of the individual spins.

\subsection{Spin eigenfunctions and representation of permutations}
\label{sec:spin_eigenfunctions}
Spin-adapted bases can be created using so-called genealogical coupling schemes based on the addition theorem of angular momentum \citep{Pauncz1979SpinEigenfunctions, Schnack2000RotationalModesMolecular, Schnalle2010CalculatingEnergySpectra}. The \emph{coupling scheme}, the order in which the individual spins are coupled, uniquely defines a spin-adapted basis ${\{\ket{\alpha, (S,M)} | \alpha\in A^{[S,M]}\}}$ where $(S,M)$ are the quantum numbers associated with the total spin and total magnetization. The internal quantum numbers are grouped in a vector $\alpha$ describing the given coupling scheme \citep{flockeSymmetricGroupApproachHeisenberg2002, Heitmann2019CombinedUseTranslational, Schnalle2010CalculatingEnergySpectra}. 
The possible values for the internal quantum numbers are given by the set $A^{[S,M]}$ whose cardinality only depends on the quantum numbers $S$, $M$ and not on the choice of the coupling scheme~\citep{Pauncz1979SpinEigenfunctions, flockeSymmetricgroupApproachStudies1997}
\begin{equation}
    \mathrm{Card}(A^{[S,M]}) = \frac{2S + 1}{N+1} \binom{N+1}{N/2-S} \, . \label{eq:cardinal_Asm}
\end{equation}
These eigenstates, also called \emph{configuration state functions} (CSF) can be expanded on the computational $\hat{\mathbf{s}}_z$-basis states basis using generalized Clebsch-Gordan coefficients~\citep{Schnalle2010CalculatingEnergySpectra}
\begin{equation}
    \ket{\alpha, (S, M)} = \sum_{\sum_i m_i = M} \braket{m_1\cdots m_N}{\alpha, (S, M)} \ket{m_1 \cdots m_N}\, . \label{eq:clebsch_gordan_expansion}
\end{equation}

Different coupling schemes have been proposed in the literature, each one with a specific feature, such as increasing the Hamiltonian sparsity or incorporating point-group symmetry~\citep{limanniCompressionSpinAdaptedMulticonfigurational2020, Li_Manni2021-tf, Schnalle2010CalculatingEnergySpectra}. 
In the remainder of this work, we will study the properties of the \textit{successive coupling schemes}, constructed recursively by positively ($\Delta S_i=+\frac{1}{2}$) or negatively ($\Delta S_i=-\frac{1}{2}$) coupling individual spins to previously spin-adapted states of $i-1$ spins. 

For illustration, we present the \textit{successive coupling scheme} for the specific example of a system of $N=8$ spins in a global singlet configuration, \ie $S=M=0$. 
Within this scheme, two non-equivalent encodings of the internal quantum numbers can be defined:
\begin{itemize}
    \item In the step encoding, the spin eigenstates are written as
        \begin{equation}
            \ket{\alpha=(\Delta S_1, \Delta S_2, \Delta S_3, \Delta S_4, \Delta S_5, \Delta S_6, \Delta S_7, \Delta S_8), (0, 0)}\, ,
        \end{equation}
        where the internal variable $\Delta S_i = +\frac{1}{2}$ if the $i$-th spin is positively coupled to the $(i-1)$ previous ones, and $\Delta S_i =-\frac{1}{2}$ otherwise. For clarity, we will use the labels $u/d$ (up/down) for the positive and negative coupling of total spins, respectively, and the labels $\alpha/\beta$ for projected spin values ${S_z=\pm \frac{1}{2}}$ of the original computational basis. 
        The angular momentum addition rules as well as the global $S=0$ total spin conservation imply the internal variables must satisfy the following conditions
        \begin{equation}
            \begin{cases} 
            \sum\limits_{i=1}^{k} \Delta{S}_i \geq 0,~\text{for}~1\leq k \leq 8\, ,\\
            \sum\limits_{i=1}^{8} \Delta{S}_i = S = 0\, . 
            \end{cases}
        \end{equation}
        
    \item In the height encoding, the spin eigenstates are written as
        \begin{equation}
            \ket{\alpha=(0, \bar{S}_1, \bar{S}_2, \bar{S}_3, \bar{S}_4, \bar{S}_5, \bar{S}_6, \bar{S}_7, \bar{S}_8), (0, 0)}\, ,
        \end{equation}
        where the internal variables are defined from the cumulative sums $\bar{S}_i = \sum_{j=1}^{i} \Delta S_j$. Due to the angular momentum addition rules and the global singlet condition, the internal variables must satisfy in this case the following conditions
        \begin{equation}
            \begin{cases} 
            | \bar{S}_{i+1} - \bar{S}_i | = 1/2,~ 1\leq i \leq 8 \, ,\\
            \bar{S}_{i} \geq 0,~\text{for}~1\leq i \leq 7,~\bar{S_8}=S=0 \, .
            \end{cases}
        \end{equation}
\end{itemize}

\begin{figure}[!ht]
    \centering
    \includegraphics[width=0.5\linewidth]{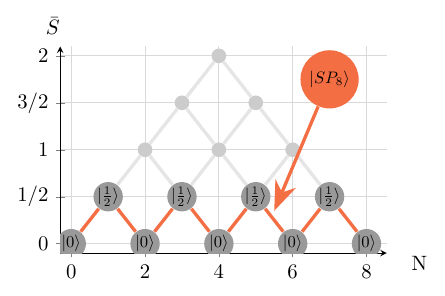}
    \caption{Yamaguchi-Kotani spin-path representation of the spin-adapted subspace for $N=8$ spins and global quantum numbers $(S,M)=(0,0)$, where each spin-path connecting the coordinates (0,0) and (8,0) corresponds to a unique total-spin eigenstate with eigenvalue $S=0$. For example, the red path represents the product of singlet pairs $\ket{SP_8}$. The encoding of this state in the height-representation, see~\cref{eq:product_spin_pairs}, is indicated by a label $\ket{s}$ in the node with coordinates $(i,s)$ corresponding to the internal quantum number $\bar{S}_i = s$.
    }
    \label{fig:yk_spinpaths}
\end{figure}

Spin-adapted states can be represented, as in~\cref{fig:yk_spinpaths}, using the Yamagouchi-Kotani spin graphs~\citep{flockeSymmetricgroupApproachStudies1997}, where they are mapped to paths connecting the initial point (0,0) to the final quantum number coordinate $(N,S)$. The height encoding of the spin-adapted states corresponds to defining the internal quantum numbers on the nodes of the path, where an eigenvalue $\bar{S}_i=s$ is associated with a spin-path passing through the node with coordinates $(i, s)$. Alternatively, the step encoding corresponds to defining the internal quantum numbers on the edges of the spin-path, where an eigenvalue $\Delta S_i = u/d$ signifies that the edge between the x-coordinates $i-1$ and $i$ is going up/down (diagonally).

The physical subspace is then composed of the states of the total Hilbert space that satisfy the constraints of the chosen encoding. For instance, the state $\ket{0,\frac{3}{2},0,\frac{1}{2},0,\frac{1}{2},0,\frac{1}{2}, 0,(0,0)}$ with the height encoding is unphysical because it does not satisfy the addition of angular momentum for a chain of spin-$1/2$s. Similarly, the state $\ket{u,d,d,u,u,d,u,d,(0, 0)}$ with the step encoding is unphysical, this time because it yields negative intermediate total spin values. An example of a physical state for spin chains with an even number of sites $N$ is the product of $N/2$ singlet-pairs (red path in \cref{fig:yk_spinpaths}), denoted as $\ket{SP_N}$,
\begin{equation}
     \ket{SP_8} = \left(\frac{\ket{\alpha \beta} - \ket{\beta \alpha}}{2}\right)^{\otimes 4} = \begin{cases}
         \text{step encoding} & \ket{u,d,u,d,u,d,u,d,(0,0)}\, ,\\
         \text{height encoding}& \ket{0, \frac{1}{2}, 0, \frac{1}{2}, 0, \frac{1}{2}, 0, \frac{1}{2}, 0, (0, 0)}\, .
     \end{cases} \label{eq:product_spin_pairs}
\end{equation}
The product of singlet-pairs state in \cref{eq:product_spin_pairs} will play an important role in the following sections. Note that the spin-adapted state $\ket{u,d,u,d,u,d,u,d,(0,0)}$ is not equivalent to the non-spin-adapted Neel-states $\ket{\alpha, \beta,\alpha, \beta,\alpha, \beta, \alpha, \beta}$ and $\ket{\beta,\alpha, \ldots}$ respectively. Different orderings of the spins in the coupling scheme correspond to different product of singlet-pairs states, and have been used in the context of the valence-bond theory to describe the low-energy spectrum \citep{White1994ResonatingValenceBond, Schwandt2014ValenceBondDistribution, Kottmann2022OptimizedLowdepthQuantum}. 


\subsection{Truncated spin-path subspaces}
\label[subsection]{truncated_subspaces}

For the total spin values considered in this work, $S=0$ and $S=1$, the number of physical states, given in \cref{eq:cardinal_Asm}, still grows exponentially with the system size. For a chain of $N$ spins in a global singlet configuration (with $N$ even), the maximum value for the intermediate total spin of a sub-chain is $\bar{S}_{max}=N/4$, as can be seen from \cref{fig:yk_spinpaths}. Recent works \citep{Vieijra2020RestrictedBoltzmannMachinesa, dobrautzCombinedUnitarySymmetric2022} have shown that, within \emph{successive coupling schemes}, the ground state of the one-dimensional antiferromagnetic Heisenberg Hamiltonian has a large overlap with the spin eigenstate that is the product of singlet-pairs $\ket{SP_N}$ (red path in \cref{fig:yk_spinpaths}), and has a large support over states with low intermediate total spin values. Based on this observation, the authors~\citep{Vieijra2020RestrictedBoltzmannMachinesa} have introduced a hierarchy of increasingly large subspaces of the total Hilbert space based on the truncation of the spin paths to a maximum value $\bar{S}_{trunc}$. The corresponding sets of valid internal variables $A^{[S,M]}_{\leq \bar{S}_{trunc}}$ are most naturally expressed in the height-encoding representation, as each internal variable directly encodes the total spin of intermediate spin chains. Below, we explicitly define the subspaces obtained with the lowest truncation levels ${\bar{S}_{trunc}=\{\frac{1}{2}, 1, \frac{3}{2}, 2\}}$ for a chain of $N$ spins in a global singlet $S=0$ configuration.

The smallest subspace one can consider corresponds to $\bar{S}_{trunc}=1/2$, and contains only one state which is the product of singlet-pairs state $\ket{SP_N}$ shown in \cref{fig:yk_spinpaths}. This state has a larger overlap with the antiferromagnetic ground state than any of the standard computational basis states, including the Neel states.

\begin{figure}[!ht]
    \centering
    \includegraphics[width=\linewidth]{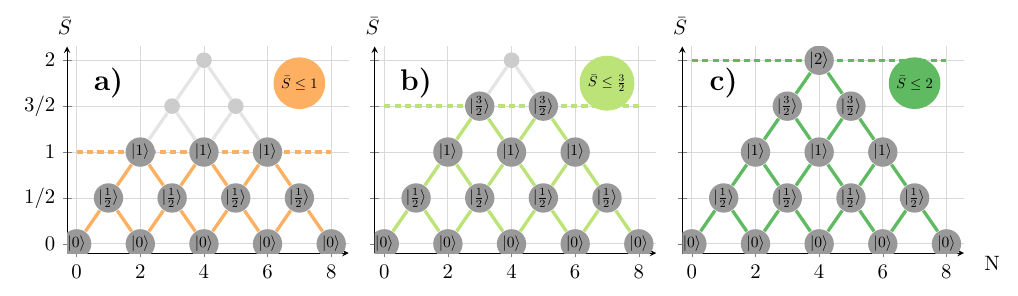}
    \caption{Yamaguchi-Kotani spin-path representation of the truncated spin-adapted subspace for $N=8,~S=0$. We depict, from left to right, the subspace representation associated with three truncation parameters $\bar{S}_{trunc} \in \{1, \frac{3}{2}, 2\}$. We note that the first subspace contains $2^{N/2-1}=2^{3}=8$ states, the second $14$ states, and the third $15$ states, which recovers the full Hilbert space. The internal local quantum states $\ket{S_i=s}$ in the height encoding are indicated for all indices $i$ by a label $\ket{s}$ in the corresponding nodes with coordinates $(i,s)$.}
    \label{fig:hilbert_space_truncation}
\end{figure}

The first subspace whose size grows exponentially with the system size and, hence, is non-trivial, corresponds to $\bar{S}_{trunc}=1$, depicted in \cref{fig:hilbert_space_truncation}(a). In this case, the only internal degrees of freedom are the intermediate total spin values for the even sites
\begin{equation}
    \bar S_{2k}\in \{0, 1\}, \quad 0<2k<N \, .
\end{equation}

In the subspace with $\bar{S}_{trunc}=3/2$, depicted in \cref{fig:hilbert_space_truncation}(b), every intermediate total spin value $\bar S_k$ on even and odd sites can take two possible values
\begin{align}
    \bar S_{2k} \in \{0, 1\} &,~0<2k<N\, , \\
    \bar S_{2k+1} \in \left\{\frac{1}{2}, \frac{3}{2}\right\} &,~0<2k+1<N \, .
\end{align} 

Lastly, we consider the $\bar{S}_{trunc}=2$ subspace, represented in \cref{fig:hilbert_space_truncation}(c). As it has been shown~\citep{Vieijra2020RestrictedBoltzmannMachinesa}, the ground state of the antiferromagnetic Heisenberg Hamiltonian for chains of up to 20 sites has a large support over this subspace. The intermediate total spin values $\bar S_k$ can take up to three possible values on even sites and up to two on odd sites
\begin{align}
    \bar S_{2k} \in \{0, 1, 2\} &,~0<2k<N \, , \\
    \bar S_{2k+1} \in \left\{\frac{1}{2}, \frac{3}{2}\right\} &,~0<2k+1<N \, .
\end{align} 

The matrix representations of the Heisenberg Hamiltonian in the truncated spin-adapted bases, labeled with the truncation level $\bar{S}_{trunc}$, are obtained as described in \cref{sec:matrix_representation}. By construction, the ground state energy estimates $(E_{gs})_{\leq \bar{S}_{trunc}}$ of the truncated subspaces satisfy the variational principle
\begin{equation}
    (E_{gs})_{exact} = (E_{gs})_{\leq \bar{S}_{max}}< \ldots < (E_{gs})_{\leq 1} < (E_{gs})_{\leq 1/2}.
    \label{eq:variational_hierarchy}
\end{equation}

\section{Representation of the Hamiltonian in spin-adapted bases}
\label[section]{sec:matrix_representation}

The most straightforward way of computing the Hamiltonian representation in spin-adapted bases is to expand the basis functions as in \cref{eq:clebsch_gordan_expansion}, and to evaluate the matrix elements in the simpler $\hat{\mathbf{s}}_z$ eigenbasis (\textit{i.e.}, the standard computational basis). Although very general, this approach is not scalable as the expansion of the spin-adapted states on the computational basis states scales exponentially with the number of spins. Here, we propose a strategy to calculate the matrix representation of the Hamiltonian in the spin-adapted basis that does not explicitly require the expansion in the original tensor-product basis. Approaches to do so can be mostly classified in two categories, the \emph{unitary group approach} (UGA)~\citep{Moshinsky1967GROUPTHEORYMANYBODY, Shavitt1977GraphTheoreticalConcepts, Shavitt1978MatrixElementEvaluation, Paldus1974GroupTheoreticalApproach, Paldus1976UnitarygroupApproachManyelectron, Paldus2021MatrixElementsUnitarya} and the \emph{symmetric group approach} (SGA)~\citep{Duch1982SymmetricGroupGraphicala, Pauncz1979SpinEigenfunctionsa, Ruedenberg1971ExpectationValuesManyFermion}. Both these schemes use properties of the $SU(2)$ group to derive rules for efficient calculation of the matrix elements. Our alternative derivation aims to preserve, within the SGA framework, the natural tensor-product structure of the calculation rules.

\subsection{Symmetric group approach for spin-adapted representations}

Following \citet{flockeSymmetricgroupApproachStudies1997} and later \citet{dobrautzEfficientFormulationFull2019}, we first remark that the $(S,M)$-adapted spin basis functions generate an irreducible representation of the symmetric group $\mathcal{S}_N$. In other words, the spin permutations $\hat{\bm{\pi}}_{i,j}\in \mathcal{S}_N$ for any ${1<i,j<N}$ do not mix different $(S,M)$-symmetry sectors. The connection between the Heisenberg Hamiltonian and the symmetric group follows from the Dirac identity
\begin{equation}
    \hat{\mathbf{s}}_i \cdot \hat{\mathbf{s}}_j = \frac{1}{2} \hat{\bm{\pi}}_{i,j} - \frac{1}{4} \hat{\mathbf{I}}\, ,
\end{equation}
which implies that the Heisenberg Hamiltonian can be expressed in terms of permutations
\begin{align}
    \hat{\mathbf{H}}_{Heisenberg} &= \frac{1}{2}\sum_{\langle i,j \rangle} J_{ij} \hat{\bm{\pi}}_{i,j} - \frac{1}{4}\sum_{i,j} J_{ij} \hat{\mathbf{I}} \, . \label{eq:heisenberg_hamiltonian_dirac}
\end{align}
When expressing \cref{eq:heisenberg_hamiltonian_dirac} in the spin-eigenbasis with quantum numbers $(S, M)$, the formal permutations $\hat{\bm{\pi}}_{i,j}$ are replaced by their matrix representation, which we denote by $\Gamma^{[S,M]}[\hat{\bm{\pi}}_{i,j}]$ (later shortened as $\Gamma[\hat{\bm{\pi}}_{i,j}]$ when the symmetry eigenvalues are unambiguous). Instead of constructing the matrix representation of each individual permutation group element, it is more convenient to first derive the representation of the elementary permutations $\Gamma[\hat{\bm{\pi}}_{i,i+1}]$, and later use the group composition to obtain all permutations as

\begin{equation}
    \Gamma[\hat{\bm{\pi}}_{i,j}] = \left(\prod_{k=i}^{j-1}\Gamma[\hat{\bm{\pi}}_{k,k+1}]\right) \Gamma[\hat{\bm{\pi}}_{j-1,j}] \left(\prod_{k=i}^{j-1}\Gamma[\hat{\bm{\pi}}_{k,k+1}]\right)^{-1} \, .
    \label{eq:group_identity}
\end{equation}

Each term in the product can be successively applied on the state of interest without explicitly calculating the large matrix products in \cref{eq:group_identity} to reduce the memory required for computing all matrix elements.
This requires only determining the action of the matrices $\Gamma[\hat{\bm{\pi}}_{i,i+1}]$ on the spin-paths, which can be evaluated from the identity~\citep{flockeSymmetricgroupApproachStudies1997}
\begin{equation}
    \Gamma[\hat{\bm{\pi}}_{i,i+1}] = \sum_s P(\bar S_{i+1}=s)\Gamma[\hat{\bm{\pi}}_{i,i+1})] = \sum_s \Gamma[(\hat{\bm{\pi}}_{i,i+1})_s] \, ,
    \label{eq:permutation_as_sum}
\end{equation}
where $P(\bar S_{i+1}=s)$ is the projector on the subspaces associated with eigenvalues of the $(i+1)$-th intermediate total spin. In the Yamaguchi-Kotani representation of symmetry subspaces, this projector selects all the paths that are going through the vertex with coordinate ${(i+1,\bar{S}_{i+1}=s)}$. The action of the projected operator $\Gamma[(\hat{\bm{\pi}}_{i,i+1})_s]$ is most easily given in terms of a set of graphical rules~\citep{flockeSymmetricgroupApproachStudies1997} acting on the spin-paths in \cref{fig:hilbert_space_truncation}
\begin{gather}
    \Gamma[(\hat{\bm{\pi}}_{i,i+1})_s] \biggl[ \upuppath \biggr] = \upuppath, \quad \Gamma[(\hat{\bm{\pi}}_{i,i+1})_s] \biggl[  \downdownpath \biggr] = \downdownpath \, , \notag\\
    \Gamma[(\hat{\bm{\pi}}_{i,i+1})_s] \biggl[ \updownpath \biggr] = -a_{s}\updownpath + b_{s} \downuppath \, , \notag \\
    \Gamma[(\hat{\bm{\pi}}_{i,i+1})_s] \biggl[ \downuppath \biggr] = a_{s}\downuppath + b_{s} \updownpath \label{eq:spin_path_rules} \, .
\end{gather}
Here, only the vertices of the spin-path corresponding to the indices $\{i-1, i, i+1\}$ and such that the rightmost node has coordinates $(i+1,s)$, corresponding to ${\bar{S}_{i+1}=s}$, are shown. Remarkably, these rules do not explicitly depend on the label $i$, except for boundary effects, which we will address in the later practical sections.
The coefficients $a_s$, $b_s$ depend on the intermediate total-spin value of the vertex $i+1$ and can be expressed as
\begin{equation}
    a_s = \frac{1}{2s+1} = \cos(\theta_s), \quad b_s = \sqrt{1-a_s^2} = \sin(\theta_s). \label{eq:as_bs_thetas}
\end{equation}
While $\Gamma[\hat{\bm{\pi}}_{i,i+1}]$ is in principle highly non-local due to the presence of the projection operator $P(\bar S_{i+1}=s)$ in \cref{eq:permutation_as_sum}, the graphical rules highlight the locality of $\Gamma[(\hat{\bm{\pi}}_{i,i+1})_s]$ in the spin-adapted basis. Indeed, for a fixed value of the intermediate total spin, this operator only modifies at most three internal quantum numbers (with indices $i-1, i, i+1$).

\subsection{SGA graphical rules in the step and height encoding}
Given an encoding of the Yamaguchi-Kotani spin-paths, the graphical rules in \cref{eq:spin_path_rules} can be translated to matrices acting on the encoding space.
In the following, we derive the representation of the graphical rules in the height and the step encoding defined in \cref{sec:spin_eigenfunctions}.

\subsubsection{Step encoding}
In the step encoding, the graphical rules in \cref{eq:spin_path_rules} can be interpreted as local operators acting on two neighboring internal quantum number vector $\ket{\Delta S_i, \Delta S_{i+1}}$
\begin{gather}
    \Gamma[(\hat{\bm{\pi}}_{i,i+1})_s] \ket{u, u}_{i, i+1} = P(\bar S_{i+1}=s)\ket{u, u}_{i, i+1}\, , \notag \\ 
    \Gamma[(\hat{\bm{\pi}}_{i,i+1})_s] \ket{d, d}_{i, i+1} = P(\bar S_{i+1}=s)\ket{d, d}_{i, i+1}\, , \notag\\
    \Gamma[(\hat{\bm{\pi}}_{i,i+1})_s] \ket{u, d}_{i, i+1} = P(\bar S_{i+1}=s)\left( -a_{s}\ket{u, d}_{i, i+1} + b_{s} \ket{d, u}_{i, i+1}\right)\, , \notag \\
    \Gamma[(\hat{\bm{\pi}}_{i,i+1})_s] \ket{d, u}_{i, i+1} = P(\bar S_{i+1}=s)\left(+b_{s}\ket{u, d}_{i, i+1} + a_{s} \ket{d, u}_{i, i+1}\right) \, . \label{eq:step_encoding_rules} 
\end{gather}
Since the intermediate total spin value is not directly accessible in the step encoding, the projector in \cref{eq:permutation_as_sum} must be explicitly expanded as
\begin{equation}
    P(\bar S_{i+1}=s) = \sum_{\Delta S_1+ \cdots +\Delta S_{i+1}=s} \ketbra{\Delta S_1, \cdots ,\Delta S_{i+1}}{\Delta S_1, \cdots ,\Delta S_{i+1}} \, , \label{eq:step_encoding_projector}
\end{equation}
acting on all the internal variables before the $(i+1)$-th index, $\Delta S_1, \dots, \Delta S_{i+1}$.

\subsubsection{Height encoding}
As opposed to the step encoding, the internal quantum numbers in the height encoding directly contain the intermediate total spin values appearing in the projectors. As a result, the projector is represented by an operator acting only on a single internal variable ${P(\bar{S}_{i+1}=s) = \ketbra{s}{s}_{i+1}}$. The set of graphical rules can be expressed in terms of operators acting only on three consecutive indices $\{i-1, i, i+1\}$
\begin{gather}
    \Gamma[(\hat{\bm{\pi}}_{i,i+1})_s] \ket{s-1, s-1/2, s} = \ket{s-1, s-1/2, s}\, , \notag\\
    \Gamma[(\hat{\bm{\pi}}_{i,i+1})_s] \ket{s+1, s+1/2, s} = \ket{s+1, s+1/2, s}\, , \notag\\
    \Gamma[(\hat{\bm{\pi}}_{i,i+1})_s] \ket{s, s+1/2, s} = -a_{s}\ket{s, s+1/2, s} + b_{s} \ket{s, s-1/2, s}\, , \notag \\
    \Gamma[(\hat{\bm{\pi}}_{i,i+1})_s] \ket{s, s-1/2, s} = +b_{s}\ket{s, s+1/2, s} + a_{s} \ket{s, s-1/2, s} \label{eq:height_encoding_rules} \, .
\end{gather}

In ground-state optimization algorithms, one can explicitly use the rules in \cref{eq:height_encoding_rules} (resp. \cref{eq:step_encoding_rules}) to compute the action of any permutation (and therefore of the full Hamiltonian) on any state of the truncated spin-adapted bases.
This allows for memory-efficient diagonalization of the full Heisenberg Hamiltonian with Lanzcos based methods and has been extensively researched in the SGA literature \citep{Shavitt1978MatrixElementEvaluation, Manley1984DirectorProgramCalculating}.

\subsection{SGA graphical rules in tensor product form}

So far, our derivation has been entirely classical. However, our ultimate goal is designing scalable quantum algorithms for measuring approximate ground-state properties for large system sizes. For applications to quantum computing, one is required to map the Hamiltonian to operators acting on the Hilbert space of qubits or qudits. Efficient quantum algorithms can only be derived if this mapping yields a sparse operator.
In the rest of this section, we embed the step-encoded and spin-adapted basis for $N$ spins in the Hilbert space of $N$ qudits (resp. $N+1$ in the height encoding) with a local dimension $d=2$ (resp. $d=\bar{S}_{max}+1$). This allows us to rewrite the SGA graphical rules as sparse operators in the basis of the $N$-qudit \emph{Pauli group} $\mathcal{P}^{2}_{N} = \{I, X, Y, Z\}^{\otimes N}$ (resp. $\mathcal{P}_{N+1}^d$).

In the step encoding, it is natural to map the step variables $\{d, u\}$ onto the qubit states $\{0, 1\}$ and to express the spin-adapted operators in terms of the qubit Pauli group $\mathcal{P}^{2}_{N} =\{I, X, Y, Z\}^{\otimes N}$. The graphical rules in \cref{eq:step_encoding_rules} can be rewritten with tensor products of Pauli group operators
\begin{equation}
    \Gamma[(\hat{\bm{\pi}}_{i,i+1})_s] = P(\bar S_{i+1}=s)\left(\frac{II + ZZ}{2} + a_s \frac{ZI - IZ}{2} + b_s \frac{XX + YY}{2}\right)_{i, i+1} \, .
    \label{eq:step_encoding_rules_pauli}
\end{equation} 

As already mentioned, the projector operator in \cref{eq:step_encoding_projector} is highly non-local, as it has support over all internal degrees of freedom preceding the $i$-th step variable. In addition to the non-locality, expressing each outer product $\ketbra{\Delta S_1, \cdots ,\Delta S_{i+1}}{\Delta S_1, \cdots ,\Delta S_{i+1}}$ in terms of the Pauli group basis, based on the the identities
\begin{equation}
    \ketbra{0}{0} = \frac{I+Z}{2},\quad
    \ketbra{0}{1} = \frac{X+iY}{2},\quad \ketbra{1}{0} = \frac{X-iY}{2},\quad \ketbra{1}{1} = \frac{I-Z}{2} \, ,
\end{equation}
results in a dense operator with, in the worst case, an exponential number of $2^{i+1}$ terms for each individual projector.

Conversely, in the height encoding of the spin-paths, the representation of the intermediate total-spin projectors remains local. This comes at the cost that the internal degrees of freedom are non-binary. As an intermediate step for enabling quantum simulations of the height-encoded Heisenberg Hamiltonian, we propose to embed the spin-adapted subspace with global quantum numbers $(S,M)$ in the Hilbert space of $(N+1)$ qudits with local dimensions $d=\bar{S}_{max}+1$. For convenience, we do not use the $(N+1)$-qudit Pauli group elements directly, but instead introduce tensor products of pseudo $I_s,X_s,Y_s,Z_s$ operators acting on the subspace $\{\ket{s-1/2}, \ket{s+1/2}\}$ of a given internal variable
\begin{align}
    Z_s &= \ketbra{s+1/2}{s+1/2} - \ketbra{s-1/2}{s-1/2} \, , \notag \\
    X_s &= \ketbra{s+1/2}{s-1/2} + \ketbra{s-1/2}{s+1/2} \, , \notag \\
    Y_s &= -i Z_s X_s \, .
\end{align}
These operators are different from the $I,X,Y,Z$ operators of the qudit Pauli group, and will serve to facilitate the mapping of the Hamiltonian to qubit operators. We rewrite the height-encoded graphical rules in \cref{eq:height_encoding_rules} as tensor-products of the $X_s$, $I_s$, and $Z_s$ Pauli matrices. This yields two terms: 1) a \emph{ZZ}-antiferromagnetic interaction that penalizes configurations with different intermediate total spin values for next-nearest neighbors, and 2) a controlled Zeeman-like interaction with an external field aligned with the direction $aZ+bX$,
\begin{align}
    \Gamma^{ZZ}[(\hat{\bm{\pi}}_{i,i+1})_s] = &\ketbra{s+1/2}{s+1/2}_{i-1}\otimes \ketbra{s-1/2}{s-1/2}_{i+1} + \notag \\
    &\ketbra{s-1/2}{s-1/2}_{i-1}\otimes \ketbra{s+1/2}{s+1/2}_{i+1} \label{eq:height_encoding_ZZ_term_v0}\\
    = & \left(\frac{I_s I_s - Z_s Z_s}{2} \right)_{i-1, i+1}\, , \label{eq:height_encoding_ZZ_term_v1}\\
    \Gamma^{aZ+bX}[(\hat{\bm{\pi}}_{i,i+1})_s] = &\ketbra{s}{s}_{i-1}\otimes \left(a_s Z_s + b_s X_s \right)_{i} \otimes \ketbra{s}{s}_{i+1} \, . \label{eq:height_encoding_aZbX_term}
\end{align}

To derive these equations, we have used the equivalence of certain projector combinations, \emph{e.g.} $\ketbra{s-1, s-1/2, s}{s-1, s-1/2, s}_{i-1,i,i+1} = \ketbra{s-1}{s-1}_{i-1} \otimes \ketbra{s}{s}_{i+1}$.
This holds because in the physical sector consecutive total spin values can only differ by $\pm 1/2$.


\subsection{Truncated subspace Hamiltonian}
\begin{figure}[ht!]
    \centering
    \includegraphics[width=0.5\linewidth]{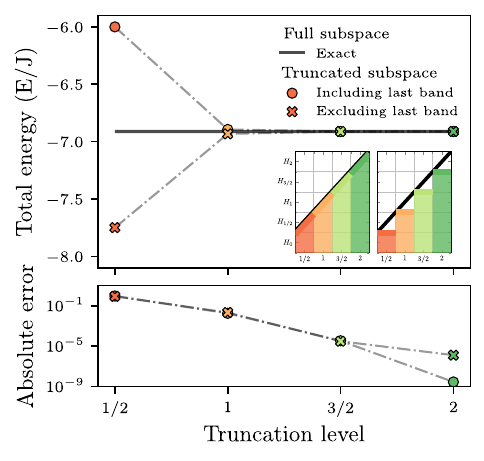}
    \caption{Convergence of the ground state energy of a 16-site Heisenberg chain for increasing truncation level from $\bar{S}_{trunc}=1/2$ to $\bar{S}_{trunc}=2$. The ground-state energy of the truncated Hamiltonian (circles) monotonously converges towards the exact energy obtained from exact diagonalization of the full Hamiltonian matrix in the spin-adapted subspace. Results obtained based on the truncated approximate Hamiltonian (crosses) are not guaranteed to converge variationally. The inset is a representation of the relation between the truncated subspaces (x-axis) and the Hamiltonian bands that contribute to these subspaces (y-axis). While the \emph{height truncation} accounts for the partial contributions of the last band, represented by the colored half-squares below the truncation line (black diagonal), the \emph{band truncation} excludes them in favor of sparsity.}
    \label{fig:figure_energy_convergence}
\end{figure}

In the previous derivation, the permutations appearing in the Heisenberg Hamiltonian are represented as sparse sums of local tensor-product operators acting on the height-encoded internal variables. To this end, we embedded the symmetry subspace into a $(N+1)$-qudit Hilbert space. However, the largest local dimension $d=\bar{S}_{max}+1$ of an internal variable in the height representation scales linearly with the system size, which is not practical on currently available quantum hardware. Instead, we adapt the subspace truncation presented in \cref{sec:spin_adapted_bases} to formulate embeddings of the truncated spin-adapted subspaces in $(N+1)$-qudit systems with a small, constant local dimension. For a truncation level $\bar{S}_{trunc}$, we can embed the truncated spin-adapted subspace in the Hilbert space,
\begin{equation}
    \mathcal{H}_N^{\bar{S}_{trunc}} = \left\{ \ket{a_0, \dots, a_N},
    \begin{array}{c}
        0\leq a_{2k} \leq \bar{S}_{trunc} \text{~integer}\\
        1/2\leq a_{2k+1} \leq \bar{S}_{trunc} \text{~half-integer}
    \end{array} \right\} \, ,
\end{equation}
with local dimension $d = \lfloor \bar{S}_{trunc} \rfloor + 1$.
Truncating the subspace preserves the locality of the operators in the height encoding and, consequently, the sparsity of the Hamiltonian representation in the Pauli group basis. 
Consider the action of the subspace truncation on the height-encoded Hamiltonian expression in the $(N+1)$-qudit Pauli basis. We indicate the truncation of the Hamiltonian representation matrix $\Gamma[\hat{\mathbf{H}}]$ as $\Gamma_{\leq \bar{S}_{trunc}}[\hat{\mathbf{H}}]$~\citep{Footnote1}. As for elementary permutations in \cref{eq:permutation_as_sum}, we decompose the Hamiltonian operator as a sum of contributions associated with intermediate total spin subspaces. A re-summation of \cref{eq:heisenberg_hamiltonian_dirac,,eq:permutation_as_sum} yields
\begin{align}
    \Gamma_{\leq \bar{S}_{trunc}}[\hat{\mathbf{H}}]  &= \frac{J}{2}\left(\sum_{i=1}^{N-1} \Gamma_{\leq \bar{S}_{trunc}}[\hat{\bm{\pi}}_{i,i+1}] - \frac{(N-1)}{2} \right) \\
                        &= \frac{J}{2}\left(\sum_{i=1}^{N-1} \sum_{s} \Gamma_{\leq \bar{S}_{trunc}}[(\hat{\bm{\pi}}_{i,i+1})_s] - \frac{(N-1)}{2} \right) \notag \\
                        &= \frac{J}{2}\left(\sum_{s} \Gamma_{\leq \bar{S}_{trunc}}[\hat{\mathbf{H}}_s] - \frac{(N-1)}{2} \right) \, , \label{eq:hamiltonian_as_sums_of_bands}
\end{align}
where in the last step of Eq.~\eqref{eq:hamiltonian_as_sums_of_bands} we have defined the $s$-band Hamiltonians $\hat{\mathbf{H}}_s$. For a given $s$ value, $\hat{\mathbf{H}}_s$ represents the reduced Heisenberg interaction within the subspace of the intermediate spin variables graphically identified by the horizontal bands of range $[s-1/2, s+1/2]$. Summing these terms up to the maximum possible intermediate total spin value $\bar{S}_{max}$ yields the exact Hamiltonian representation in the truncated subspace. The terms of the Hamiltonian corresponding to $s>\bar{S}_{trunc}$ in \cref{eq:hamiltonian_as_sums_of_bands} all contain projectors on high intermediate total spin values and all evaluate to 0 in the truncated subspace. Therefore, we can express the truncation of the subspace as a truncation of the band summation appearing in \cref{eq:hamiltonian_as_sums_of_bands} as
\begin{equation}
    \Gamma_{\leq \bar{S}_{trunc}}[\hat{\mathbf{H}}] = \frac{J}{2}\left(\sum_{s=0}^{\bar{S}_{trunc}} \Gamma_{\leq \bar{S}_{trunc}}[\hat{\mathbf{H}}_s] - \frac{(N-1)}{2} \right) \, . \label{eq:sparse_qudit_hamiltonian}
\end{equation}
Particular care must be paid when accounting for the last band, corresponding to $s=\bar{S}_{trunc}$, as it couples states of the truncated subspace to the rest of the spin-adapted subspace. For instance, the $\Gamma^{ZZ}$ terms of this band all evaluate to 0 after the subspace truncation, as they are composed of products of two projectors, one on the intermediate total spin value $s=\bar{S}_{trunc}-1/2$, below the truncation level, and one on $s=\bar{S}_{trunc}+1/2$, above the truncation level. Conversely, the terms $\Gamma^{aZ+bX}[(\hat{\bm{\pi}}_{i,i+1})_s]$ of this band contain coupling terms between the truncated subspace and the rest of the symmetry subspace, which evaluate to 0 after the subspace truncation, but also coupling terms within the truncated subspace, which result in
\begin{equation}
    \Gamma_{\leq \bar{S}_{trunc}}[H_{s=\bar{S}_{trunc}}] = \sum_{i=1}^{N-1} \ketbra{s}{s}_{i-1}\otimes \left(\frac{a_s}{2} \ketbra{s-1/2}{s-1/2} \right)_{i} \otimes \ketbra{s}{s}_{i+1} \, .
\end{equation}
While including these contributions ensures that the so-constructed Hamiltonian yields the exact \emph{height-truncated} subspace Hamiltonian matrix, their decompositions on the basis of tensor products of the $I_s, X_s, Y_s, Z_s$ matrices require 8 terms (2 for each projector in the tensor product). This renders the Pauli representation of the truncated Hamiltonian significantly less sparse. In the following, we define a slightly different hierarchy of Hamiltonians, $\Gamma_{<\bar{S}_{trunc}}[\hat{\mathbf{H}}]$, where the last band is excluded. We define these \emph{band-truncated} Hamiltonians, as opposed to the \emph{height-truncated} matrix $\Gamma_{\leq \bar{S}_{trunc}}[\hat{\mathbf{H}}]$. This sequence of Hamiltonians still converges to the exact Hamiltonian when $\bar{S}_{max}$ bands are included. However, the corresponding ground-state energies may violate \cref{eq:variational_hierarchy}, i.e., they may not be variational. We show the convergence of the ground state energy for a 16-site Heisenberg chain and for increasing truncation values $\bar{S}_{trunc} \in \{1/2,1,3/2,2\}$ in \cref{fig:figure_energy_convergence}. While the ground state energy sequence for the height-truncated Hamiltonians $\Gamma_{\leq \bar{S}_{trunc}}[\hat{\mathbf{H}}]$ (blue crosses) is monotonously decreasing because of the variational principle, that of the band-truncated Hamiltonians $\Gamma_{< \bar{S}_{trunc}}[\hat{\mathbf{H}}]$ does not (green crosses, only converges in absolute difference).

\section{Qubit mapping of truncated height-encoded Hamiltonians}
\label[section]{sec:qudit_to_qubit_mapping}

We now turn to the question of mapping the truncated, spin-adapted representation of the Heisenberg Hamiltonian derived above onto a register of qubits for applications in qubit-based quantum computing. We will focus our attention on the sparsity and locality of the Pauli representation of the derived operators. As already discussed in \cref{truncated_subspaces}, the low-energy spectrum of the isotropic and antiferromagnetic Heisenberg Hamiltonian can be well approximated within the truncated spin-adapted subspaces already for small truncation levels. In this section, we construct qubit encodings of the \emph{band-truncated} Hamiltonians for truncation levels $\bar{S}_{trunc}\in \{1/2, 1, 3/2\}$. The calculation for $\bar{S}_{trunc}=2$ is detailed in \cref{sec:hamiltonian_and_unitary_0020}. We will denote as $\Gamma[\hat{\mathbf{H}}]$ the Hamiltonian represented in the $(N+1)$-qudit Hilbert space, as opposed to $\overline{\Gamma}[\hat{\mathbf{H}}]$ when accounting for the qudit-to-qubit encoding.

\subsection{Truncation level \texorpdfstring{$\bar{S}_{trunc}=1/2$}{Strunc=1/2}}
For $\bar{S}_{trunc}=1/2$, the Hamiltonian is a single matrix element given by the energy of the tensor product of singlet pairs $\ket{SP_N}$. From  \cref{eq:sparse_qudit_hamiltonian}, it can be evaluated as
\begin{align}
    \Gamma_{< 1/2}[\hat{\mathbf{H}}]
    &= \frac{J}{2}\left( \frac{-N}{2} - \frac{(N-1)}{2}  \right) \ketbra{SP_N}{SP_N} \notag \\
    &= \frac{J}{2}\left(-N + \frac{1}{2}\right) \ketbra{SP_N}{SP_N} \, .
\end{align}

\subsection{Truncation level \texorpdfstring{$\bar{S}_{trunc}=1$}{Strunc=1}}
Within the lowest non-trivial truncation scheme, all odd intermediate total spin values are fixed to the value $\bar{S}_{2k+1}=1/2$, and all even values can be represented by a qubit encoding the values $\{0, 1\}$. This allows us to map the general expression of the $(N+1)$-qudit Hamiltonian in \cref{eq:sparse_qudit_hamiltonian} onto an $N/2$-qubit Hamiltonian. 
The total Hamiltonian in this subspace is the sum of the two bands from \cref{eq:sparse_qudit_hamiltonian} for $s=0$ and $s=1/2$. As a first step, we simplify the trivial projectors on the odd sites and obtain an $N/2$-qudit Hamiltonian
\begin{align}
    \Gamma_{< 1}[\hat{\mathbf{H}}_{0}] &= \sum_{2i=0}^{N-1} (-a_0) \ketbra{0}{0}_{2i} \otimes \ketbra{0}{0}_{2i+2} \, ,\\
    \Gamma_{< 1}[\hat{\mathbf{H}}_{1/2}] &= \sum_{2i+1=1}^{N-1} \Gamma^{ZZ}_{<1}[(\hat{\bm{\pi}}_{2i+1,2i+2})_{1/2}] +  \sum_{2i=2}^{N-1}  \Gamma^{aZ+bX}_{<1}[(\hat{\bm{\pi}}_{2i,2i+1})_{1/2}] \, .
\end{align}

The contribution from the band $s=1$ that we discarded for the sake of sparsity is 
\begin{equation}
    \Gamma_{\leq 1}[\hat{\mathbf{H}}_{1}] = \sum_{2i=0}^{N-1} (a_1) \ketbra{1}{1}_{2i}\otimes \ketbra{1}{1}_{2i+2} \, .
\end{equation}

The even and odd permutations in $\hat{\mathbf{H}}_{1/2}$ are given by \cref{eq:height_encoding_ZZ_term_v1,eq:height_encoding_aZbX_term}. After the qubit encoding, the pseudo Pauli matrices $Z_{1/2}, X_{1/2}$ for the even indices are mapped to the qubit Pauli Matrices $Z, X$, yielding
\begin{align}
    \overline{\Gamma}^{aZ+bX}_{<1}[(\hat{\bm{\pi}}_{2i,2i+1})_{1/2}] &= (a_{1/2} Z + b_{1/2} X)_{2i} \, ,\\
    \overline{\Gamma}^{ZZ}_{<1}[(\hat{\bm{\pi}}_{2i+1,2i+2})_{1/2}] &= \left(\frac{II - ZZ}{2}\right)_{2i, 2i+2} \, .
\end{align}

The qubit encoding of $\hat{\mathbf{H}}_{0}$ and $\hat{\mathbf{H}}_{1/2}$ are respectively given by a diagonal and a \emph{tilted-field Ising Hamiltonian} acting on $N/2$ qubits, \textit{i.e.}
\begin{align}
    \overline{\Gamma}_{< 1}[\hat{\mathbf{H}}_{0}] &= \sum_{2i=0}^{N-1} -a_0 \left(II+IZ+ZI+ZZ\right)_{2i,2i+2} \, ,\\
    \overline{\Gamma}_{< 1}[\hat{\mathbf{H}}_{1/2}] &= \sum_{2i=0}^{N-1} \left(\frac{II - ZZ}{2}\right)_{2i, 2i+2} + (a_{1/2} Z + b_{1/2} X)_{2i} \, .
\end{align}

\subsection{Truncation level \texorpdfstring{$\bar{S}_{trunc}=3/2$}{Strunc=3/2}}
\label[section]{truncation_level_S32}
The total Hamiltonian in the $\bar{S}_{trunc}=3/2$ subspace is the sum of the first three bands from \cref{eq:sparse_qudit_hamiltonian}, corresponding to $s=0$, $s=1/2$ and $s=1$
\begin{align}
    \Gamma_{< 3/2}[\hat{\mathbf{H}}_{0}] &= \sum_{2i=0}^{N-1} \ketbra{0}{0}_{2i}\otimes (-a_0 \ketbra{1/2}{1/2})_{2i+1} \otimes \ketbra{0}{0}_{2i+2} \, ,\\
    \Gamma_{< 3/2}[\hat{\mathbf{H}}_{1/2}] &= \sum_{2i+1=1}^{N-1} \Gamma^{ZZ}_{<3/2}[(\hat{\bm{\pi}}_{2i+1,2i+2})_{1/2}] +  \sum_{2i=2}^{N-1} \Gamma^{aZ+bX}_{<3/2}[(\hat{\bm{\pi}}_{2i,2i+1})_{1/2}] \, ,\\
    \Gamma_{< 3/2}[\hat{\mathbf{H}}_{1}] &= \sum_{2i=2}^{N-1} \Gamma^{ZZ}_{<3/2}[(\hat{\bm{\pi}}_{2i,2i+1})_{1}] +  \sum_{2i+1=1}^{N-1} \Gamma^{aZ+bX}_{<3/2}[(\hat{\bm{\pi}}_{2i+1,2i+2})_{1}] \, .
\end{align}
The discarded contribution from the band $s=3/2$ reads
\begin{equation}
    \Gamma_{\leq 3/2}[\hat{\mathbf{H}}_{3/2}] = \sum_{2i=2}^{N-2} \ketbra{3/2}{3/2}_{2i-1} \otimes (a_{3/2} \ketbra{1}{1})_{2i} \otimes \ketbra{3/2}{3/2}_{2i+1} \, .
\end{equation}

The local Hilbert space comprises the states $\{\ket{0}, \ket{1}\}$ for even indices and $\{\ket{1/2}, \ket{3/2}\}$ for odd indices. In both cases, we can encode the local Hilbert space at each index with a single qubit. The sparse $N$-qudit Hamiltonian in \eqref{eq:sparse_qudit_hamiltonian} is mapped to a $N$-qubit Hamiltonian by mapping the pseudo Pauli matrices $Z_{1/2}, X_{1/2}$ (resp. $Z_{1}, X_{1}$) to the conventional qubit Pauli matrices $Z, X$ for even (resp. odd) indices,
\begin{align}
    \overline{\Gamma}^{aZ+bX}_{<3/2}[(\hat{\bm{\pi}}_{2i,2i+1})_{1/2}]  &= \ketbra{0}{0}_{2i-1} \otimes (a_{1/2} Z + b_{1/2} X)_{2i} \otimes \ketbra{0}{0}_{2i+1} \, ,\\
    \overline{\Gamma}^{ZZ}_{<3/2}[(\hat{\bm{\pi}}_{2i+1,2i+2})_{1/2}]   &= \left(\frac{II - ZZ}{2}\right)_{2i, 2i+2} \, ,\\
    \overline{\Gamma}^{aZ+bX}_{<3/2}[(\hat{\bm{\pi}}_{2i+1,2i+2})_{1}]  &= \ketbra{1}{1}_{2i} \otimes(a_{1} Z + b_{1} X)_{2i+1} \otimes \ketbra{1}{1}_{2i+2} \, ,\\
    \overline{\Gamma}^{ZZ}_{<3/2}[(\hat{\bm{\pi}}_{2i,2i+1})_{1}]       &= \left(\frac{II - ZZ}{2}\right)_{2i-1, 2i+1} \, .
\end{align}

The qubit Hamiltonians for the zeroth, first, and second bands are given by a diagonal matrix and two \emph{controlled-tilted-field Ising Hamiltonians} respectively,
\begin{align}
    \overline{\Gamma}_{< 3/2}[\hat{\mathbf{H}}_{0}] &= \sum_{2i=0}^{N-1} -a_0 \ketbra{000}{000}_{2i,2i+1,2i+2} \, ,\\
    \overline{\Gamma}_{< 3/2}[\hat{\mathbf{H}}_{1/2}] &= \sum_{2i=0}^{N-1} \left(\frac{II - ZZ}{2}\right)_{2i, 2i+2} + \ketbra{0}{0}_{2i-1} \otimes (a_{1/2} Z + b_{1/2} X)_{2i} \otimes \ketbra{0}{0}_{2i+1} \, ,\\
    \overline{\Gamma}_{< 3/2}[\hat{\mathbf{H}}_{1}] &= \sum_{2i=0}^{N-1} \left(\frac{II - ZZ}{2}\right)_{2i-1, 2i+1} + \ketbra{1}{1}_{2i} \otimes (a_{1} Z + b_{1} X)_{2i+1} \otimes \ketbra{1}{1}_{2i+2} \, .
\end{align}

\subsection{Spin-path boundary conditions}
\label[section]{sec:spin_path_boundary_conditions}
The band Hamiltonian construction presented so far is independent from the global symmetry eigenvalues $(S,M)$ of the system. As can be seen from \cref{fig:hilbert_space_truncation}, the spin-paths are subject to boundary conditions, starting at the origin $(0,0)$ and finishing at the coordinates~$(N,S)$. While the Heisenberg Hamiltonian can be expressed as a sum of Hamiltonian independently of these boundaries, one advantage of the expression in \cref{eq:sparse_qudit_hamiltonian} is the natural accounting of the spin-subspace truncation.
For large systems and small total spin eigenvalues, the global symmetry eigenvalues amount to boundary conditions on top of the Hamiltonian band description in the bulk.

We mainly study spin-chains with an even number of sites $N$ in a global singlet $S=0$ configuration. In this case, the intermediate total spin values at the boundaries are fixed to $\bar{S}_0=\bar{S}_{N}=0$. From the addition of angular momentum rules, it also follows that $\bar{S}_{1}=\bar{S}_{N-1}=1/2$. One advantage of the spin-adapted basis is that the global triplet subspace can be targeted through a simple change of boundary conditions. It is sufficient to set the last internal variable to $\bar{S}_{N}=1$, and consider an additional degree of freedom $\bar{S}_{N-1}\in\{1/2,3/2\}$. Although we do not consider the application of the proposed construction to total spin values greater than $S>1$, our framework can be extended also to these cases.

In most of our previous derivations, the Hamiltonian bands are derived without accounting for the boundary conditions. We found it convenient to introduce the boundary restrictions in a second step, both for the Hamiltonians derived in \cref{sec:qudit_to_qubit_mapping} and for the corresponding time-evolution unitaries derived in \cref{sec:time_evolution_truncated_hamiltonians}. As an example, let us detail the calculation of the matrix elements of the permutation $(1,2)$ in the subspace $\bar{S}_{trunc}=1$. We first derive the \emph{band}-permutations using \cref{eq:height_encoding_aZbX_term,eq:height_encoding_ZZ_term_v1}
\begin{align}
    \Gamma_{<1}[\hat{\bm{\pi}}_{1,2}] &= \sum_{s=0}^{1/2} \Gamma_{<1}[(\hat{\bm{\pi}}_{1,2})_s] \notag \\ 
                                &= -a_0 \ketbra{0}{0}_{0}\otimes \ketbra{0}{0}_{2} + \left(\frac{II - Z_{1/2}Z_{1/2}}{2}\right)_{0, 2} \, .
\end{align}
To apply the boundary conditions at the origin of the spin-path, $\bar{S}_0=0$, $\bar{S}_{1}=1/2$, we multiply this expression by a projector $P_{bc} = \ketbra{0, 1/2}{0, 1/2}_{0,1}$, and discard the corresponding (non-dynamical) eigenvalues from the vector of internal variables $\alpha$
\begin{align}
    \Gamma^{bc}_{<1}[\hat{\bm{\pi}}_{1,2}] 
                    &=  \Gamma_{<1}[\hat{\bm{\pi}}_{1,2}] P_{bc} \notag \\
                    &= -a_0 \ketbra{0}{0}_{2} +\ketbra{1}{1}_{2} \, . \label{eq:fixing_boudaries}
\end{align}

When mapping the Hamiltonian onto the qubit register, the number of relevant degrees of freedom required to describe each of the subspaces depends on the total-spin-symmetry eigenvalue. More precisely, the spin-path boundary conditions effectively reduce of the number of qubits required for the representing the Hamiltonian within the truncated subspaces.

\section{Time evolution under truncated Hamiltonians}
\label[section]{sec:time_evolution_truncated_hamiltonians}

Based on the theory defined in the previous section, we derive an adiabatic state-preparation schedule for preparing the ground state of the Heisenberg Hamiltonian~\citep{Ciavarella2023SimulatingHeisenbergInteractions, Ciavarella2023StatePreparationHeisenberg} in a spin-adapted basis. 
Practical implementations of adiabatic state preparation are limited by the ability to perform precise real-time dynamical simulations under a time-dependent Hamiltonian. In the usual $\hat{\mathbf{s}}_z$-computational basis, real-time dynamics of spin Hamiltonians have already been subject to extensive research both with classical~\citep{Langer2011RealtimeEnergyDynamics, Rodriguez2022FarfromequilibriumUniversalityTwodimensional} and quantum computing \citep{CoelloPerez2022QuantumStatePreparation, Yoshioka2024DiagonalizationLargeManybodya, Ma2011QuantumSimulationWavefunction}. Exact quantum-dynamics simulations are limited to small system sizes because of the high computational cost related to the evaluation of the unitary time-evolution operator (for time-independent Hamiltonians)
\begin{equation}
    \hat{\mathbf{U}}_{\mathrm{exact}}(T) = \mathcal{T}\exp\left(-i \int_0^T \hat{\mathbf{H}}(t') dt' \right) 
    =
    \exp(-i T \hat{\mathbf{H}}) \, .
\end{equation}
This high computational cost can be tamed down by resorting to Trotter-Suzuki formulas, where the time-evolution operator is expanded into products of small evolution steps with bounded approximation errors. For a Hamiltonian composed of two non-commuting parts $\hat{\mathbf{H}}=\hat{\mathbf{H}}_A+\hat{\mathbf{H}}_B$, the first- and second-order Trotter-Suzuki formulas read
\begin{align}
    \hat{\mathbf{U}}_{O_1}(T) &\approx \left(\exp(-i \frac{T}{N_L} \hat{\mathbf{H}}_A) \exp(-i \frac{T}{N_L} \hat{\mathbf{H}}_B) \right)^{N_L} \, , \\
    \hat{\mathbf{U}}_{O_2}(T) &\approx \left(\exp(-i \frac{T}{2 N_L} \hat{\mathbf{H}}_A) \exp(-i \frac{T}{N_L} \hat{\mathbf{H}}_B) \exp(-i \frac{T}{2 N_L} \hat{\mathbf{H}}_A \right)^{N_L} \, .
\end{align}
Given a choice of representation for the operators, or equivalently of a computational basis, these formulas can be implemented as quantum circuits with $N_L$ repeated layers. Expectation values of general observables on time-evolved states can be estimated through measurements on the quantum circuit. Contrary to the ideal time dynamics, which preserve all the symmetries of the ideal Hamiltonian, the Trotter decomposition involves an algorithmic approximation, due to the non-commutativity of the individual terms. This approximation can lead to symmetry contamination. Here, we propose a variant of the Heisenberg time evolution, where the Trotterized real-time dynamics are formulated in the truncated spin-adapted bases presented in \cref{sec:spin_adapted_bases}. Our sequence of approximation unitaries is not affected by spin contamination and converges to the Trotterized real-time dynamics in the limit of large truncated subspaces.

\subsection{Non spin-adapted time evolution}

\begin{figure}[ht!]
    \centering
    \includegraphics[width=\linewidth]{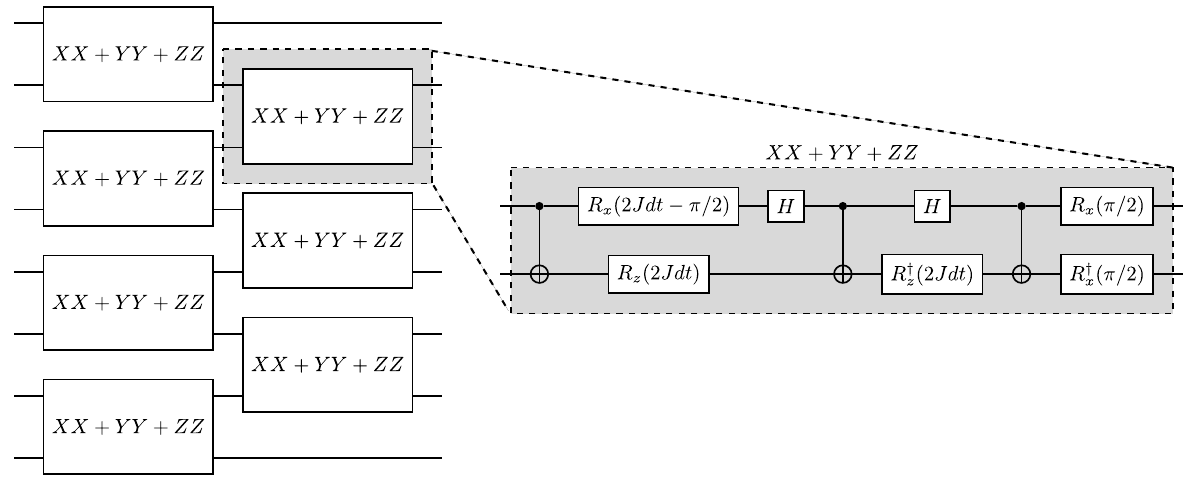}
    \caption{Left) Even- and odd-layer decomposition of one Trotter step of the Heisenberg unitary evolution of the unitary $(\overline{U}^{\hat{\mathbf{s}}_z})_{O_1}$ for the first order Trotter formulas. Right) 3-CX optimal decomposition of the 2-qubit Heisenberg interaction.}
    \label{fig:heisenberg_xxyzz}
\end{figure}

The nearest-neighbor 1D Heisenberg model is usually expressed in the computational $\hat{\mathbf{s}}_z$-basis, where each individual spin is encoded in a single qubit. The corresponding qubit Hamiltonian, in terms of the Pauli group elements, reads
\begin{align}
    \overline{\Gamma}^{\hat{\mathbf{s}}_z}[\hat{\mathbf{H}}] &= J \sum_{i} \overline{\Gamma}^{\hat{\mathbf{s}}_z}[\hat{\mathbf{s}}_i \cdot \hat{\mathbf{s}}_{i+1}] = J \sum_{i} \left(\frac{XX + YY + ZZ}{4}\right)_{i,i+1}\\
    &= J \sum_{2i} \left(\frac{XX + YY + ZZ}{4}\right)_{2i,2i+1} + J \sum_{2i+1} \left(\frac{XX + YY + ZZ}{4}\right)_{2i+1,2i+2}. \label{eq:H_xxyyzz}
\end{align}
The Trotter-Suzuki formula associated with this Hamiltonian requires splitting $\hat{\mathbf{H}}$ into an even and and odd part, $\hat{\mathbf{H}}=\hat{\mathbf{H}}_{even}+\hat{\mathbf{H}}_{odd}$, to simulate the evolution for a small time $dt$. This results in two sets of two-qubit interactions that can be applied in parallel for all even and all odd pairs of spins. The circuit implementation of the unitary $(\overline{U}^{\hat{\mathbf{s}}_z})_{O_1}$ for the first order Trotter formulas is shown in \cref{fig:heisenberg_xxyzz}. The two-qubit interaction is further decomposed into a universal basis of gates, composed of single qubit gates and CX gates. Using the layout of \cref{fig:heisenberg_xxyzz}, this requires seven single qubit gates and three two-qubit CX gates~\citep{Vidal2004UniversalQuantumCircuit}.

\subsection{Approximated Trotterized dynamics in truncated spin-adapted bases}
As already discussed in \cref{truncated_subspaces}, the low-energy spectrum of the Heisenberg Hamiltonian has a large overlap with the truncated spin-adapted basis states. Consequently, the real time-evolution of a low-energy initial state is mainly driven by the interactions that are included in the truncated Hamiltonians for small truncation levels. Based on the expression of the qubit Hamiltonians in the truncated subspaces in \cref{sec:qudit_to_qubit_mapping}, we construct the unitary circuits simulating the corresponding approximated Trotterized time evolution. Moreover, we show that our construction quickly converges toward the Trotterized dynamics under the full Hamiltonian. For clarity, we denote the Hamiltonian representations on the qubit Pauli group in the $s<\bar{S}_{trunc}$ truncated spin-adapted subspace as $\overline{\Gamma}^{csf}_{s<\bar{S}_{trunc}}[\hat{\mathbf{H}}]$.

\subsubsection{Elementary blocks for building approximated Trotter unitaries}
\begin{figure}[ht!]
    \centering
    \includegraphics[width=1\linewidth]{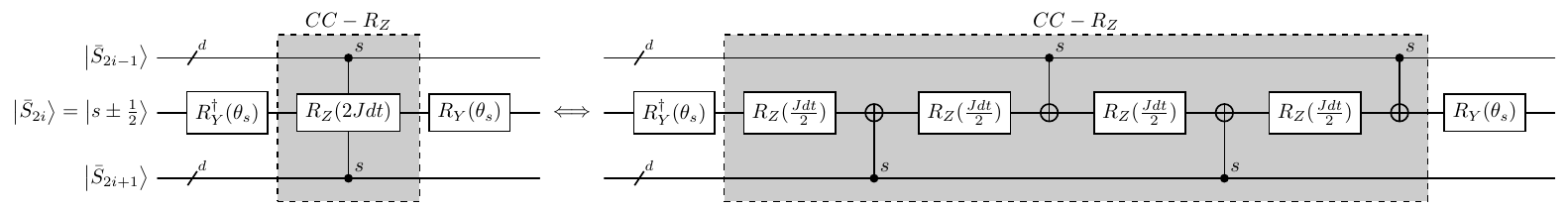}
    \includegraphics[width=0.5\linewidth]{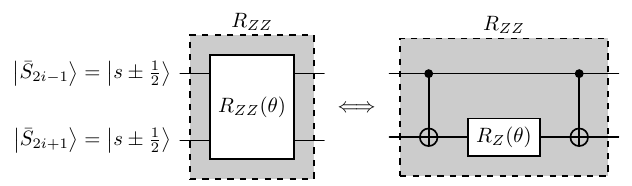}
    \caption{Qudit unitaries implementing the time-evolution under the elementary interactions $\Gamma^{aZ+bX}$ and $\Gamma^{ZZ}$ appearing in the spin-adapted Hamiltonian. Top) Decomposition of the double-controlled rotation into single qubit rotations and $CX$ entangling gates, where it is understood that the control on a qudit register corresponds to the qudit projector $\ketbra{s}{s}$. We use two qudit registers for encoding the height variables at sites $2i-1$ and $2i+1$, and a single qubit register for encoding the states $\ket{\bar{S}_{2i} = s\pm \frac{1}{2}}$ at the index $2i$ in \cref{eq:ccrx}. Bottom) Same decomposition for the next-nearest-neighbor Ising interaction. We use two qubit registers to encode the states $\ket{\bar{S}_{2i-1}} = \ket{s\pm \frac{1}{2}}$ and $\ket{\bar{S}_{2i+1}} = \ket{s\pm \frac{1}{2}}$ in \cref{eq:ii_zz}.}
    \label{fig:tikz_circuit_CCRZ_RZIZ}
\end{figure}

\label[section]{sec:elementary_blocks}
Within the smallest subspace with $\bar{S}_{trunc}=1/2$, the time-dynamics of the state $\ket{SP_N}$ simply consist of a global phase, and all time-dependent observables are constant. Starting from the subspace with $\bar{S}_{trunc}=1$, the Hamiltonian bands are composed of two types of interactions $\Gamma^{aZ+bX}$ and $\Gamma^{ZZ}$. Before mapping the Hamiltonian to the qubit space, the first contribution takes the form of a three-qudit interaction between neighboring degrees of freedom. The unitary that implements the corresponding time-evolution reads
\begin{align}
    \exp\biggl[-iJdt \Gamma^{aZ+bX}&[(\hat{\bm{\pi}}_{2i, 2i+1})_s]\biggr] \notag \\
    &= \exp\biggl[-i Jdt \ketbra{s}{s} \otimes (a_{s} Z_s + b_{s} X_s)\otimes \ketbra{s}{s}\biggr]_{2i-1, 2i, 2i+1}  \, , \notag \\
    &= R_{Y_s}^\dag(\theta_s)_{2i} \exp\left[-iJdt \ketbra{s}{s} \otimes Z_s \otimes \ketbra{s}{s} \right]_{2i-1, 2i, 2i+1} R_{Y_s}(\theta_s)_{2i} \, , \label{eq:ccrx}
\end{align}
and represents a rotation in the subspace $\ket{\bar{S}_{2i}=s\pm \frac{1}{2}}$ of the qudit at the index $2i$, controlled by the state of the qudits at the location $2i-1$ and $2i+1$. We report in \cref{fig:tikz_circuit_CCRZ_RZIZ} the circuits that implement this evolution in a three-qudit register. 

The second interaction term takes the form of a $ZZ$-Ising interaction in the subspace $\{\ket{\bar{S}_{2i}=s\pm \frac{1}{2}}, \ket{\bar{S}_{2i}=s\pm \frac{1}{2}}\}$ of next-nearest-neighboring degrees of freedom
\begin{align}
    \exp\biggl[-iJdt \Gamma^{ZZ}[(\hat{\bm{\pi}}_{2i, 2i+1})_s]\biggr] &= \exp\biggl[-i Jdt \left(\frac{II-Z_sZ_s}{2}\right)\biggr]_{2i-1,2i+1}  \, . \label{eq:ii_zz}
\end{align}
We show in \cref{fig:tikz_circuit_CCRZ_RZIZ} how to expand these terms (up to a global phase) on the universal gate set composed of single qubit rotation and the $CX$ entangling gate.

\subsubsection{Trotter circuits for the truncated Hamiltonian in the \texorpdfstring{$\bar{S}_{trunc}=3/2$}{Strunc=3/2} subspace}

In our construction, the Hamiltonian evolution is decomposed into even and odd layers as in \cref{eq:H_xxyyzz}, each further decomposed into the contributing Hamiltonian bands. Bands $\hat{\mathbf{H}}_{s}$ are composed of $\Gamma^{aZ+bX}$ interactions on odd (resp. even) indices and $\Gamma^{ZZ}$ interactions on even (resp. odd) indices, depending on the parity of $s$. Notably, the particular arrangement of the CX gate in \cref{fig:tikz_circuit_CCRZ_RZIZ} allows for a parallel implementation of all interactions acting on even (resp. odd) indices within a single band. While the elementary blocks in \cref{fig:tikz_circuit_CCRZ_RZIZ} formally act on qudit registers, the truncation of the symmetry subspace to $\bar{S}_{trunc}=3/2$ yields a local dimension $d=2$ (see \cref{truncation_level_S32}). As a result, the qudit unitaries for the elementary $\Gamma^{aZ+bX}$ and $\Gamma^{ZZ}$ interactions become qubit unitaries. 

\begin{figure}[!ht]
    \centering
    \includegraphics[width=\linewidth]{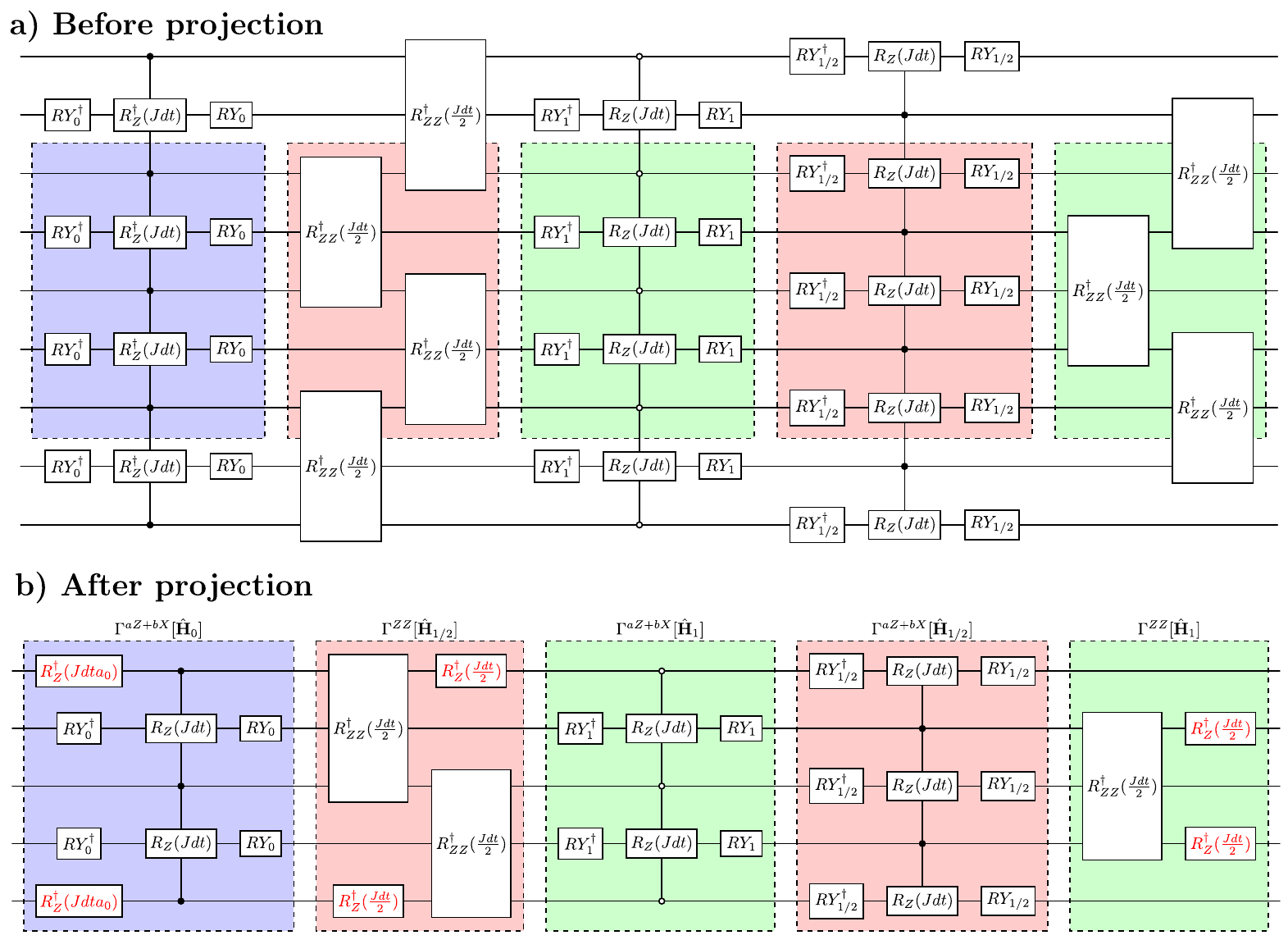}
    \caption{a)~Quantum circuit implementation of the unitary evolution in the $\bar{S}<3/2$ subspace for a time step $dt$. The total unitary is split into even and odd parts, each decomposed into commuting Hamiltonian band unitaries for $s=0$, $s=1/2$ and $s=1$, respectively highlighted in blue, red, and green. The first two and the last two qubits must all be set to the state $\ket{0}$ for a global singlet configuration, which will require projecting out all qubits not highlighted in the circuit. For readability, we use the notation $RY_{s} = R_Y(\theta_s)$ with $\theta_s$ defined in \cref{eq:as_bs_thetas}. b)~Final circuit after projection of the boundary conditions corresponding to the global singlet, fixing the first two and last two qubit states to $\ket{0}$ ($\overline{P}_{bc}=\ketbra{00}{00}_{0,1}\otimes\ketbra{00}{00}_{N-1,N}$). Note that three-qubit unitary blocks $R_{ZZ}$ are in fact two-qubit unitaries acting only on the first and the last qubit of the block.}
    \label{fig:tikz_circuit_0015_old_new_labels}
\end{figure}

In \cref{fig:tikz_circuit_0015_old_new_labels}(a) we show a quantum circuit implementation of the Trotter circuit for the truncated Hamiltonian within the $\bar{S}<3/2$ subspace, for a chain of $N=8$ sites in a global singlet configuration $S=0$. The three bands $s=0$, $s=1/2$, and $s=1$ contribute to each odd layer and are implemented in three unitaries colored in blue, red and green, respectively. Conversely, only the bands $s=1/2$, and $s=1$ contribute to the even permutations and are implemented by the unitaries colored in red and green, respectively. At this stage, each internal quantum number of the spin chain is encoded into one qubit, giving a total unitary acting on $N+1=9$ qubits.

After implementing the Hamiltonian bands as layers in a quantum circuit, we must pay special attention to the terminal qubits of the register. 
For an even number of sites in a global singlet configuration, the first and last two qubits are projected in the $\ket{0}$ state, see \cref{sec:spin_path_boundary_conditions}. This projection can be directly performed on the circuit of \cref{fig:tikz_circuit_0015_old_new_labels}(a), by simplifying the circuit gates. For example, the $CX$ gates with a fixed, non-dynamical control qubit in the zero (resp. one) state, reduces to an identity (resp. to an $X$-gate) on the target qubit. As a result, one obtains the quantum circuit acting on $N+1-4=5$ qubits shown in~\cref{fig:tikz_circuit_0015_old_new_labels}(b).

\subsubsection{Trotter layers for other truncated symmetry subspaces}

\begin{figure}[!ht]
    \centering
    \includegraphics[width=1\linewidth]{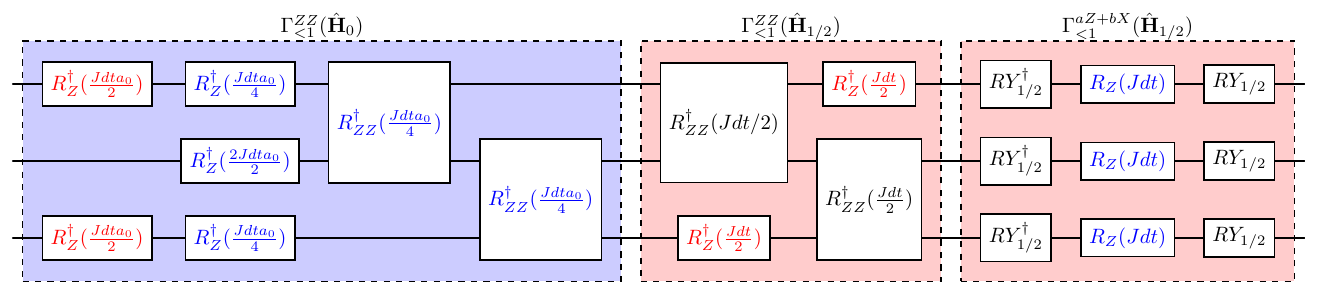}
    \caption{Circuit implementing the unitary evolution in the $\bar{S}\leq 1$ subspace for a time $dt$ under the Hamiltonian for the bands $s=0$ (blue background) and $s=1/2$ (red background). Note that this circuit could be further simplified, by combining gates with common generators when commutation rules allow it.}
    \label{fig:tikz_circuit_0010}
\end{figure}

We obtain the Trotter circuit for the smaller truncated subspace corresponding to ${\bar{S}_{trunc}=1}$ similarly to what has been described above. By definition, this subspace is obtained by fixing the odd qubits to $\bar{S}_{2k+1}=1/2$. In the circuits provided in \cref{fig:tikz_circuit_0015_old_new_labels}, this corresponds to projecting out the qubit lines at even indices onto the $\ket{0}$ state, and yields a quantum circuit acting on a register of $N/2 - 1 = 3$ qubits as shown in \cref{fig:tikz_circuit_0010}.

\subsection{Comparison of spin-adapted and \textit{standard} time-evolution}

We show here that our proposed sequence of Trotterized circuits in the truncated spin-adapted basis approaches the exact time-dynamics (within the error bounds of the Trotter approximation) even for small truncated subspaces. We use the Trotterized time-dynamics in the \textit{standard} $\hat{\mathbf{s}}_z$-computational basis as the reference for the Trotter errors, and discuss the convergence properties of our sequence of approximation unitaries with respect to the exact time-dynamics.

The generators of the exact unitary time-evolution $U_{exact}$, the permutation operators that define the Hamiltonian, commute with the symmetry operators. As such, the ideal dynamics of an initial state diagonalizing the total spin operator remains in the initial symmetry subspace, even if this state was prepared in a non spin-adapted basis. For a fair comparison, the initial state in the desired symmetry sector must then be chosen such that it can be efficiently prepared both in the spin-adapted and non spin-adapted basis. While the two representations become equivalent at $\bar{S}_{trunc}=\bar{S}_{max}$ for any initial state, we expect to see effects of the truncation for the subspaces with $\bar{S}_{trunc}\in \{1/2, 1, 3/2, 2\}$. These effects are the smallest for low-energy initial states with large support in the truncated subspaces.

\begin{figure}[ht!]
    \centering
    \includegraphics[width=0.7\linewidth]{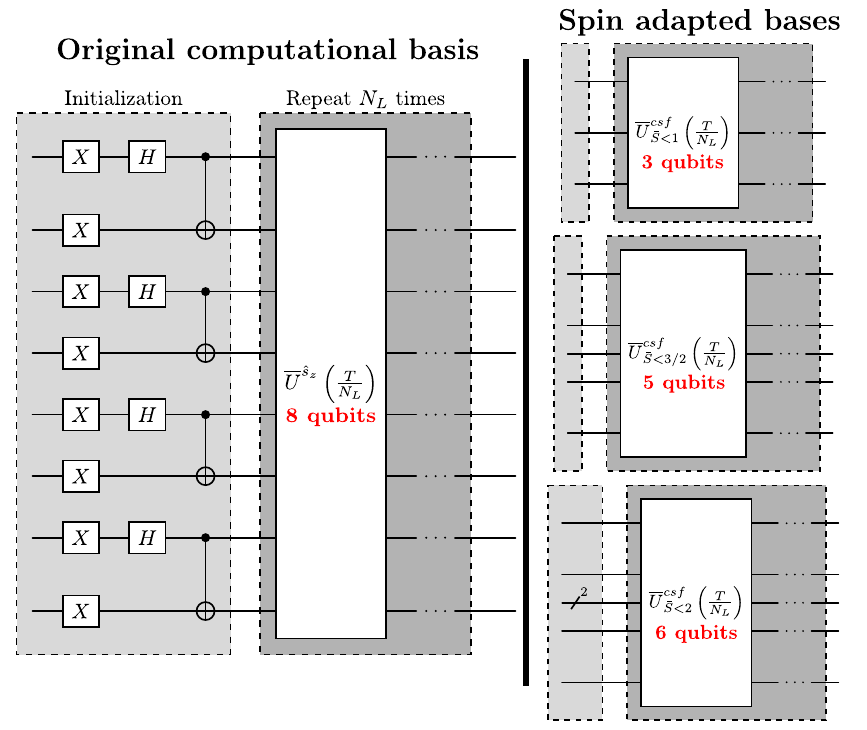}
    \caption{Left) Example Trotter circuit for the time evolution of the initial state $\ket{SP_8}$ in the standard $\hat{\mathbf{s}}_z$-basis. Right) Approximation unitaries $\overline{U}^{csf}_{s<\bar{S}_{trunc}}$ for the Trotterized dynamics of the same state, in spin-adapted bases truncated at $\bar{S}_{trunc}\in \{1,3/2,2\}$, requiring respectively 3, 5, and 6 qubits. The trivial unitary in the subspace $\bar{S}_{trunc}=1/2$ is not included in this figure.}
    \label{fig:tikz_circuit_SD_CSF}
\end{figure}

One natural choice for low-energy, easy to prepare, global singlet configuration is the tensor product of singlet pairs $\ket{SP_N}$. This state is a one of the computational basis state in the spin-adapted representation, and can be prepared in the non-spin-adapted basis using Bell-type circuits. The quantum circuit for the state initialization and the Trotter time-evolution in the standard representation, as well as the truncated spin-adapted counterparts for $\bar{S}_{trunc}\in \{1/2, 1, 3/2, 2\}$, are represented in \cref{fig:tikz_circuit_SD_CSF}.

\begin{figure}[ht!]
    \centering
    \includegraphics[width=\linewidth]{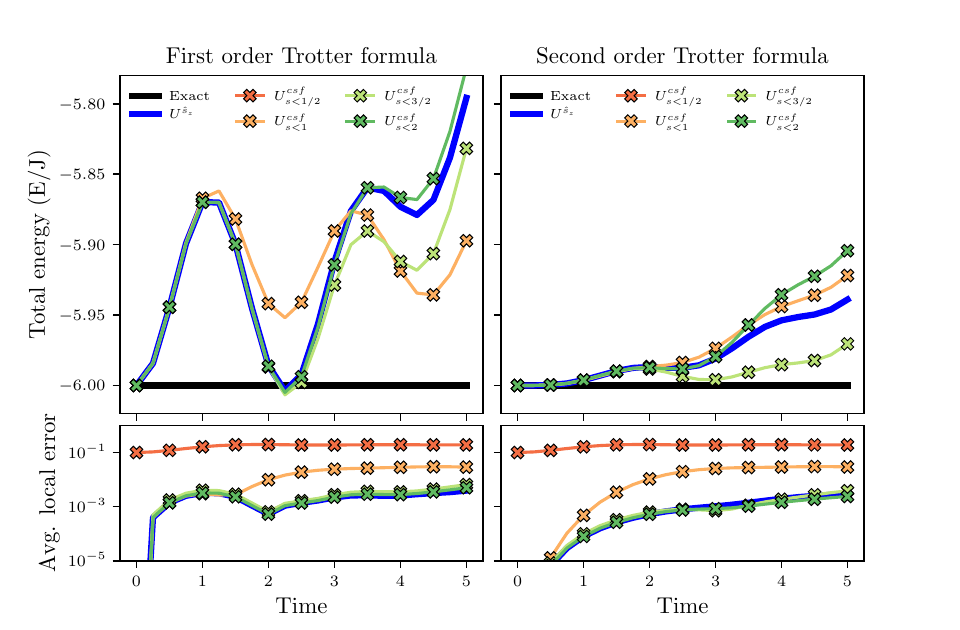}
    \caption{Upper row: Time evolution of the total energy of the state $\ket{SP_{16}}$ for a 16-site Heisenberg chain. Lower row: Average absolute error of the time-evolved local energies as a function of time for all unitaries considered in this work. The time-evolved observables are evaluated for the non-spin adapted unitary circuit (blue line) as well as our approximation unitaries in truncated spin-adapted subspaces (red, orange, light green, dark green crosses), for both the first- and second-order Trotter formulas with $N_L=10$ layer repetitions. The total energy in the subspace $\bar{S}_{trunc}<1/2$ is equal to the constant value $E_{tot}/J=-4.5$ and is not included in the window of the upper plots.}
    \label{fig:FIG10_combined_plot_total_energy_avg_error}
\end{figure}

We evaluate the accuracy of the approximated time-evolved state on a set of spin-independent observables that can be easily estimated on the above-defined quantum circuits. In \cref{fig:FIG10_combined_plot_total_energy_avg_error}, we represent the total energy of a chain of $N=16$ spins as a function of time for the different approximation circuits, for both the first and second order Trotter formulas with $N_L=10$ repetitions of the Trotter layer. The total energy is, in principle, a constant of motion of the exact time dynamics. Its conservation is broken in the simulations because of Trotter-induced algorithmic error and truncation-induced representation error. Due to the accumulation of approximation errors, the total energy evaluated with Trotter formulas for the non-spin-adapted (blue continuous line) and spin-adapted Hamiltonians (red, orange, light green, dark green lines with crosses) drift away from the expected total energy (black line). On top of the Trotter approximation, the band-truncation in the spin-adapted representation results in approximated dynamics with four level of accuracy (red, orange, light green and dark green for the truncation levels $\bar{S}_{trunc}=1/2,1,3/2$, and $2$ respectively). The first order Trotter formula shows a smooth convergence of the approximated spin-adapted time-evolutions toward the reference Trotter dynamics for increasing truncation levels $\bar{S}_{trunc}$. We see a similar trend for the second order formula, apart from the fact that the second best approximation (light green) appears to perform better than the best approximation (dark green). For this reason, we also show the average absolute error on the local energies for the two Trotter approximations in \cref{fig:FIG10_combined_plot_total_energy_avg_error}. We observe a systematic improvement of the average absolute error for increasingly large truncated subspaces for both the first- and second-order Trotter formulas.


\newpage
\section{Adiabatic ground state preparation}
\label[section]{sec:adiabatic_ground_state_preparation}

In \cref{sec:time_evolution_truncated_hamiltonians}, we showed that the time dynamics of low-energy initial states are well captured by the approximate time-evolution unitaries in the truncated subspaces, even for the smaller truncation values $\bar{S}_{trunc}\in \{1/2, 1, 3/2, 2\}$.
We make use of the proposed time-evolution unitaries in the context of adiabatic ground state preparation, and present a novel adiabatic schedule for preparing approximations of the ground state wavefunction in the spin-adapted basis using the decomposition of the Heisenberg Hamiltonian of \cref{eq:hamiltonian_as_sums_of_bands}. Similar to the approximated time-evolution circuits presented in \cref{sec:time_evolution_truncated_hamiltonians}, we mainly consider the smaller truncated subspaces. Note that, within these subspaces, the Trotter circuit complexity is reduced or comparable to that obtained in the \textit{standard} computational basis.

\subsection{Adiabatic schedule}

The tensor product of singlet pairs state used as initial point in the previous section is the non-degenerate ground state of the zeroth-band Hamiltonian $H_0$. We use this fact to build ground state approximations with an adiabatic schedule for each truncated subspace presented in \cref{sec:qudit_to_qubit_mapping}. Our adiabatic state-preparation schedules rely on a time dependent Hamiltonian that interpolates between a constant Hamiltonian $\bar{\Gamma}^{csf}_{s<\bar{S}_{trunc}}[\hat{\mathbf{H}}_0]$, with a trivial ground state $\ket{SP_N}$, and the full subspace Hamiltonian $\bar{\Gamma}^{csf}_{s<\bar{S}_{trunc}}[\hat{\mathbf{H}}]$, whose ground state we aim at preparing. Due to the band-by-band structure of the Hamiltonian in the spin-adapted basis, the following schedule implements the desired interpolation for any choice of subspace
\begin{equation}
    \bar{\Gamma}^{csf}_{s<\bar{S}_{trunc}}[\hat{\mathbf{H}}(t)] = \bar{\Gamma}^{csf}_{s<\bar{S}_{trunc}} \left[\hat{\mathbf{H}}_{0} + \sum_{s'<s} \lambda_{s'}(t) \hat{\mathbf{H}}_{s'}\right] \quad \text{where}\quad \begin{cases}
        \lambda_{s'}(0) = 0 \, ,\\
        \lambda_{s'}(T) = 1 \, .
    \end{cases}
    \label{eq:Trotter_adiabatic_schedule}
\end{equation}
Advanced step functions with smooth derivatives have been proposed in the literature to reduce the non-adiabatic effects of the schedule. For simplicity, we choose to work here with linear step functions $\lambda_{s'}(t) =t/T$ where $T$ is the total duration of the schedule.

\subsection{Low depth and low qubit count ground state approximation}
\label{sec:low_depth_gs_approximation}
We find that the smaller truncation subspace with $\bar{S}<1$ is a good candidate for preparation of ground state approximations of the Heisenberg Hamiltonian on quantum hardware. Indeed, the requirements in quantum resources are reduced compared to the \textit{standard} representation of the Heisenberg Hamiltonian, both in terms of qubit count and circuit depth. Quantitatively, the number of qubits required is only half of the number of spins in the chain, and the single Trotter layer can be reduced to single qubit rotations and two-qubit $ZZ$-interactions. The corresponding time-dependent Hamiltonian is
\begin{equation}
    \bar{\Gamma}^{csf}_{s<1}[\hat{\mathbf{H}}(t)] = \bar{\Gamma}^{csf}_{s<1} \left[\hat{\mathbf{H}}_{0} + \lambda_{1/2}(t) \hat{\mathbf{H}}_{1/2}\right] \quad \text{where}\quad \begin{cases}
        \lambda_{1/2}(0) = 0 \, ,\\
        \lambda_{1/2}(T) = 1 \, .
    \end{cases}
    \label{eq:Trotter_adiabatic_schedule_0010}
\end{equation}

which can be implemented through the Trotter layers depicted in \cref{fig:tikz_circuit_0010}, with an additional time-dependency that is accounted for by replacing the constant Trotter time-step $dt$ with the time- and band-dependent time step $dt \times \lambda_{s}(t)$.

\begin{figure}[!ht]
    \centering
    \includegraphics[width=1\linewidth]{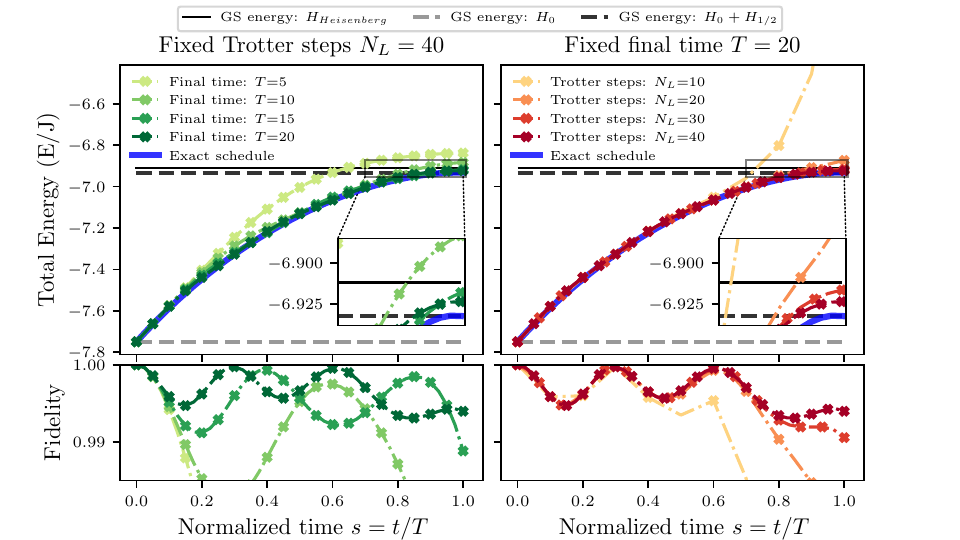}
    \caption{Adiabatic schedule in the $\bar{S}<1$ truncation subspace for: left) a fixed number of Trotter layers $N_L=40$ and an increasing adiabatic schedule duration $T$, right) a fixed schedule duration $T=20$ and an increasing number of Trotter layers. The horizontal lines represent the exact ground state energies of the Heisenberg chain for 16 sites (black line) and the exact ground state energy of the initial (resp. target) Hamiltonian of the adiabatic schedule (dashed gray, resp. black, line). The instantaneous fidelity of \cref{eq:instantaneous_fidelity}, represented in the lower panel, is calculated with respect to the exact adiabatic schedule (blue line).}
    \label{fig:0010_H0_LH12_duo_energies_fidelities_combined}
\end{figure}

In \cref{fig:0010_H0_LH12_duo_energies_fidelities_combined}, we show a realization of an adiabatic transition between the ground state of the initial Hamiltonian $\bar{\Gamma}^{csf}_{s<1}[\hat{\mathbf{H}}_0]$ (dashed gray line), and the ground state of the final Hamiltonian in the subspace $\bar{\Gamma}^{csf}_{s<1}[\hat{\mathbf{H}}_0+\hat{\mathbf{H}}_{1/2}]$ (dashed black horizontal line). Note that, at this truncation level, the ground state in the truncated subspace (dashed black horizontal line) is still visibly distinct from the exact ground state energy in the full spin-adapted subspace (full black horizontal line). To evaluate the accuracy of our proposed schedule, we first measure the energy during the Trotter schedule and compare it with the exact time dependent ground state energy of \cref{eq:Trotter_adiabatic_schedule_0010} (blue line). For the system size used in this example ($N=16$), we also evaluate exactly the instantaneous fidelity of the trotterized quantum circuit wavefunction with respect to the exact adiabatic schedule
\begin{equation}
    F(T, N_L) = \lvert \bra{SP_{16}} U_{\mathrm{exact}}^\dag(T) U_{O_2,N_L}(T) \ket{SP_{16}} \rvert \, . \label{eq:instantaneous_fidelity}
\end{equation}
We report the time-dependent energies and instantaneous fidelities as a function of time for various durations of the Trotter schedule $T \in [5, 10, 15, 20]$ (green curves with crosses), and for various numbers of Trotter layers $N_L \in [10, 20, 30, 40]$ (red curves with crosses).

\subsection{Ground state approximation in the \texorpdfstring{$\bar{S}<3/2$}{S<3/2} subspace}

Applying the same schedules in the $\bar{S}<3/2$ truncated spin-adapted subspaces leads to better ground-state energy estimates, due to the convergence of the subspace hierarchy toward the full spin-adapted subspace. As already discussed in \cref{sec:time_evolution_truncated_hamiltonians}, the unitaries implementing the corresponding time-evolution are more complex, as both the qubit count and the circuit depth depend on the truncation level. We simulate the adiabatic schedule of \cref{eq:Trotter_adiabatic_schedule} in the subspace with $\bar{S}<3/2$, with the time-dependent Hamiltonian
\begin{equation}
    \bar{\Gamma}^{csf}_{s<3/2}[\hat{\mathbf{H}}(t)] = \bar{\Gamma}^{csf}_{s<3/2} \left[\hat{\mathbf{H}}_{0} + \lambda_{1/2}(t) \hat{\mathbf{H}}_{1/2} +  \lambda_{1}(t) \hat{\mathbf{H}}_{1}\right] \, \text{where}\, \begin{cases}
        \lambda_{1/2}(0) = \lambda_{1}(0) = 0 \, ,\\
        \lambda_{1/2}(T) = \lambda_{1}(T) = 1 \, .
    \end{cases}
    \label{eq:Trotter_adiabatic_schedule_0015}
\end{equation}

\begin{figure}[!ht]
    \centering
    \includegraphics[width=1\linewidth]{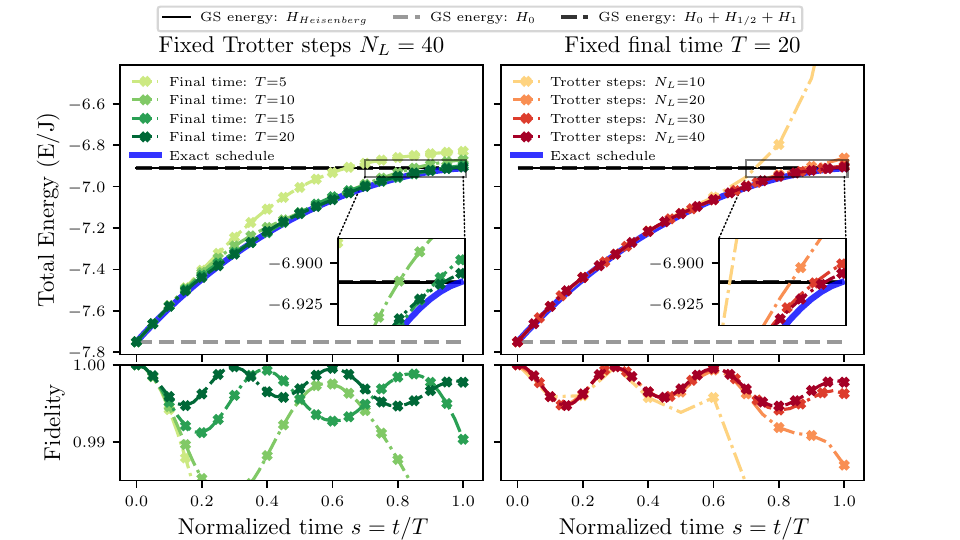}
    \caption{Adiabatic schedule in the subspace with truncation $\bar{S}<3/2$ for: left) a fixed number of Trotter layers $N_L=40$ and an increasing adiabatic schedule duration $T$, right) a fixed schedule duration $T=20$ and an increasing number of Trotter layers. The horizontal lines represent the exact ground state energies of the Heisenberg chain for 16 sites (black line) and the exact ground state energy of the initial (resp. target) Hamiltonian of the adiabatic schedule (dashed gray, resp. black, line). Note that the black and dashed black horizontal lines are indistinguishable on this scale, as they differ in absolute value by $10^{-5} J$. The instantaneous fidelity of \cref{eq:instantaneous_fidelity}, represented in the lower panel, is calculated with respect to the exact adiabatic schedule (blue line).}
    \label{fig:0015_H0_LH12H1_duo_energies_fidelities_combined}
\end{figure}

The results of our simulations are reported in \cref{fig:0015_H0_LH12H1_duo_energies_fidelities_combined} and should be compared with the results in the smaller subspace $\bar{S}<1$ in \cref{fig:0010_H0_LH12_duo_energies_fidelities_combined}. In this larger subspace, the target Hamiltonian $\bar{\Gamma}^{csf}_{s<1} \left[\hat{\mathbf{H}}_{0} + \hat{\mathbf{H}}_{1/2}\right]$ of \cref{sec:low_depth_gs_approximation} has been replaced with $\bar{\Gamma}^{csf}_{s<3/2} \left[\hat{\mathbf{H}}_{0} + \hat{\mathbf{H}}_{1/2} + \hat{\mathbf{H}}_{1}\right]$, which reduces the approximation error due to the subspace truncation alone to approximately $10^{-5} J$, see \cref{fig:figure_energy_convergence}. At the scales of the figure, the ground state of the target Hamiltonian (full black line) is indistinguishable from the exact result with the full Hamiltonian (dashed black line), and the only error comes from the Trotter approximations and from the non-adiabaticities of the time-dependent schedule. Similarly to \cref{fig:0010_H0_LH12_duo_energies_fidelities_combined}, these errors can be reduced by increasing the duration time of the Trotter schedule (green curves in the left column) and by increasing the number of Trotter steps (red curves in the right column). 
Numerically, the final instantaneous fidelities at the end of the two Trotterized schedules reach the remarkably high values of $99.26\%$ in the subspace $\bar{S}<1$, and $99.68\%$ in the subspace $\bar{S}<3/2$. This is obtained with a simple linear ramp which does not include any optimization. Additionally, the accumulated Trotter errors and non-adiabaticity effects do not increase in the larger truncated subspace, but remain of the same order of magnitude. 

\section{Conclusions}
\label[section]{sec:conclusion}
In this work, we propose a novel approach for implementing approximate quantum simulations in total-spin eigenbases, designed to produce shallow circuits that are well-suited to the capabilities of current quantum hardware.
We show that the symmetric group approach offers a natural foundation for constructing spin-adapted quantum Hamiltonians and corresponding unitaries.
This is accomplished through an encoding of spin eigenstates in the \emph{succesive coupling scheme}, combined with a truncation of the permissible values for intermediate total spin variables. This truncation is justified by the remarkably fast convergence of the ground-state wavefunctions in the truncated subspaces toward the ground-state wavefunction in the complete spin-adapted subspace.
Among the two natural encodings of the spin eigenstates, we find that the height encoding is better suited than the step encoding to quantum-computing applications, as it preserves the locality of the quantum operators.
Combining the height encoding and the truncation of height variables, we derive qubit Hamiltonians that are local and sparse, and well describe the low-energy spectrum of the Heisenberg model.
Based on our newly-introduced hierarchy of Hamiltonians, we derive adiabatic schedules between an easily prepared spin-adapted initial state and the exact ground state of the model.
We demonstrate, based on numerical simulations, that these adiabatic schedules can be used to prepare approximate, spin-adapted eigenstates of the one-dimensional Hamiltonian with relatively shallow circuits. The accuracy of our approximation is set by the number of Trotter steps (as in any adiabatic state preparation) and by the choice of the truncated subspace.

Compared to quantum algorithms that aim to preserve quantum numbers, our adiabatic ground-state preparation approach is formulated directly in the total-spin eigenbasis restricted to the relevant spin-sector.
Consequently, the resulting wavefunction can be sampled in the eigenbasis of the total spin operator $\hat{\mathbf{S}}^2$, rather than the eigenbasis of the projected spin $\hat{\mathbf{s}}_z$.
Conventional QMC simulations~\citep{limanniResolutionLowEnergyStates2021} have demonstrated that the ground-state wave function of strongly correlated spin Hamiltonians, as well as electronic structure Hamiltonians, can be significantly compressed (in the $L_1$ norm) through this basis transformation. While the quantum Schur transformation theoretically enables mapping between these representations, the corresponding unitary circuit is impractical for qubit-based quantum hardware. In contrast, our approach achieves similar results by employing a hierarchy of sparse, local, and hardware-efficient qubit Hamiltonians.



After demonstrating that our proposed protocol can be implemented using relatively shallow circuits and a simple model of the Heisenberg Hamiltonian, our future focus will be on deploying non-integrable spin Hamiltonians and ultimately electronic structure Hamiltonians on actual quantum hardware to evaluate its scalability.
By incorporating scalable error mitigation strategies~\cite{Temme2017, Cai2023, Takagi2022, Lolur2023, VanDenBerg2023_PEC}, specifically tailored to spin-adapted operators, we are optimistic that our methods can be extended to enable utility-scale quantum simulations~\cite{Kim2023EvidenceUtilityQuantum}.


In future work, we will also explore whether the $L_1$-compression of the ground-state wavefunction in the spin-adapted basis can be exploited to accelerate quantum simulations further.
We anticipate that this compression will reduce the sampling overhead required for estimating observables on states expressed in a spin-adapted basis, using techniques such as classical shadow tomography
~\cite{GarcaPrez2021_LearningToMeasure,Huang2022_Shadows,Fischer2024_DualFrame-POVM}. Furthermore, the reduction of the sampling overhead would make it appealing to combine our proposed method with the recently proposed sample-based quantum diagonalization (SQD)~\citep{Robledo-Moreno2024ChemistryExactSolutions} algorithm.
This method calculates the ground-state wave-function of a many-body quantum Hamiltonian by classically diagonalizing it within a subspace of configurations, which is generated by collecting samples from a quantum circuit that approximates the ground-state wave function.
Employing an approximate, spin-adapted wave function (prepared with our proposed adiabatic schedule) can enhance the SQD efficiency as the wavefunction is more compact, and therefore the convergence with the number of samples is expected to accelerate.


\section{Acknowledgments}
This research was supported by the NCCR MARVEL (grant number 205602), a National Centre of Competence in Research, and the project RESQUE (Rethinking Quantum Simulations in the Quantum Utility Era, grant number 20QU-1\_225229), both funded by the Swiss National Science Foundation. 
W.D. acknowledges funding from the European Union’s Horizon Europe research and innovation programme under the Marie Skłodowska-Curie grant agreement no. 101062864.

\bibliography{biblio}

\begin{thebibliography}{114}
\bibitem[{Foo()}]{Footnote1}
\bibinfo{note}{Note that the height-truncated spin-adapted subspaces are not proper representations of the symmetry group $S_N$, as the group product is broken by the height-truncation. In particular, truncated permutation matrices in the height-representation do not necessarily square to identity.}
\end{thebibliography}

\appendix
\section{Hamiltonian and unitary evolution in the subspace \texorpdfstring{$\bar{S}_{trunc}=2$}{Strunc=2}}
\label[section]{sec:hamiltonian_and_unitary_0020}

In this section we derive the Hamiltonian and the corresponding Trotter circuit in the subspace $\bar{S}_{trunc}=2$, finding additional simplifications that make it implementable on quantum hardware with a heavy-hex topology~\cite{Chamberland2020_HeavyHex}.

We start from the general expression of the embedded spin-adapted Hamiltonian in the Hilbert space of $(N+1)$-qudits with local dimension $d=\bar{S}_{trunc}+1=3$ \cref{eq:sparse_qudit_hamiltonian}

\begin{equation}
    \Gamma_{\leq 2}[\hat{\mathbf{H}}] = \frac{J}{2}\left(\sum_{s=0}^{2} \Gamma_{\leq 2}[\hat{\mathbf{H}}_s] - \frac{(N-1)}{2} \right) \, ,
\end{equation}
and then list the first contributions
\begin{align}
    \Gamma_{\leq 2}[\hat{\mathbf{H}}_0] &= \sum_{2i+1} \ketbra{0}{0}_{2i} \otimes (-a_0\ketbra{0}{0}_{2i+1}) \otimes \ketbra{0}{0}_{2i+2}\, , \\
    \Gamma_{\leq 2}[\hat{\mathbf{H}}_{1/2}] &= \sum_{2i+1} \left( \ketbra{0}{0}_{2i} \otimes \ketbra{1}{1}_{2i+2} + \ketbra{1}{1}_{2i} \otimes \ketbra{0}{0}_{2i+2} \right) \notag \\
    &+ \sum_{2i} \left( \ketbra{1/2}{1/2}_{2i-1} \otimes (a_{1/2}Z_{1/2}+b_{1/2}Z_{1/2})_{2i} \otimes \ketbra{1/2}{1/2}_{2i+1} \right)\, , \\
    \Gamma_{\leq 2}[\hat{\mathbf{H}}_{1}] &= \sum_{2i+1} \left( \ketbra{1}{1}_{2i} \otimes (a_{1}Z_{1}+b_{1}Z_{1})_{2i+1} \otimes \ketbra{1}{1}_{2i+2} \right) \notag \\
    &+ \sum_{2i} \left( \ketbra{1/2}{1/2}_{2i-1} \otimes \ketbra{3/2}{3/2}_{2i+1} + \ketbra{3/2}{3/2}_{2i-1} \otimes \ketbra{1/2}{1/2}_{2i+1} \right)\, ,\\
    \Gamma_{\leq 2}[\hat{\mathbf{H}}_{3/2}] &= \sum_{2i+1} \left( \ketbra{1}{1}_{2i} \otimes \ketbra{2}{2}_{2i+2} + \ketbra{2}{2}_{2i} \otimes \ketbra{1}{1}_{2i+2} \right) \notag \\
    &+ \sum_{2i} \left( \ketbra{3/2}{3/2}_{2i-1} \otimes (a_{3/2}Z_{3/2}+b_{3/2}X_{3/2}) \otimes \ketbra{3/2}{3/2}_{2i+1} \right) \, .
\end{align}

We can naturally encode the two possible odd-index values $\{1/2,3/2\}$ in a single qubit as done for the other subspaces.
However, we now require two qubits to encode the three possible even-site values $\{0, 1, 2\}$.
We find that simplifications arise when using a Gray-code instead of the binary encoding for the qudit-to-qubit mapping
\begin{equation}
    \ket{0}_i \rightarrow \ket{00}_{ext(i),i}, \ket{1}_i \rightarrow \ket{01}_{ext(i),i}, \ket{2}_i \rightarrow \ket{11}_{ext(i),i}, \ket{3}_i \rightarrow \ket{10}_{ext(i),i} \, , \label{eq:graycode}
\end{equation}
where the two registers required to encode the $i$-th qudit are labelled with the indices $(ext(i),i)$, with $ext$ a mapping to an external register of additional qubits.
The above expressions for the Hamiltonian band operators require only two and three body interactions (for the $\Gamma^{zz}$ and the $\Gamma^{aZ+bX}$ interactions respectively), highlighting the locality of these operators when acting on qudits. However, these three-body qudit interactions become 5-body interactions after mapping them to the qubit space in \cref{eq:graycode}, which can be challenging to implement efficiently on quantum circuits. To reduce the complexity of these terms, we leverage the freedom one has to add matrix elements in the unphysical part of the Hilbert space, without changing the physical spectrum of the Hamiltonian. We introduce the symbol $(\equiv)$ for the equivalence of the operators in the physical subspace only, and simplify the Hamiltonian as follows:

\begin{align}
    \overline{\Gamma}_{\leq 2}[\hat{\mathbf{H}}_0] &= \sum_{2i+1} \ketbra{00}{00}_{ext(2i), 2i} \otimes (-a_0\ketbra{0}{0}_{2i+1}) \otimes \ketbra{00}{00}_{ext(2i+2), 2i+2} \notag \\
    &\textcolor{black}{\equiv \sum_{2i+1} \ketbra{0}{0}_{2i} \otimes (-a_0 Z)_{2i+1} \otimes \ketbra{0}{0}_{2i+2} }\, ,
\end{align}
\begin{align}
    \overline{\Gamma}_{\leq 2}[\hat{\mathbf{H}}_{1/2}] &= \sum_{2i+1} \ketbra{00}{00}_{ext(2i), ext(2i+2)} \otimes \left( \frac{II-ZZ}{2} \right)_{2i, 2i+2} \notag \\
    &+ \sum_{2i} \left( \ketbra{0}{0}_{2i-1} \otimes \ketbra{0}{0}_{ext(2i)} \otimes (a_{1/2}Z+b_{1/2}X)_{2i} \otimes \ketbra{0}{0}_{2i+1} \right)  \notag \\
    &\textcolor{black}{\equiv \sum_{2i+1} \left( \frac{II-ZZ}{2} \right)_{2i,2i+2} \notag  }\\
    &+ \textcolor{black}{\sum_{2i} \left( \ketbra{0}{0}_{2i-1} \otimes (a_{1/2}Z+b_{1/2}X)_{2i} \otimes \ketbra{0}{0}_{2i+1} \right) }\, ,
\end{align}
\begin{align}
    \overline{\Gamma}_{\leq 2}[\hat{\mathbf{H}}_{1}] &= \sum_{2i+1} \left( \ketbra{01}{01}_{ext(2i),2i} \otimes (a_{1}Z+b_{1}X)_{2i+1} \otimes \ketbra{01}{01}_{ext(2i+2),2i+2} \right) \notag \\
    &+ \sum_{2i} \left( \frac{II-ZZ}{2}\right)_{2i-1, 2i+1 }  \notag\\
    &\textcolor{black}{\equiv \sum_{2i+1} \left( \frac{II-ZZ}{2}\right)_{ext(2i),2i} \otimes (a_{1}Z+b_{1}X)_{2i+1} \otimes \left(\frac{II-ZZ}{2} \right)_{ext(2i+2),2i+2} \notag }\\
    &+ \textcolor{black}{\sum_{2i} \left( \frac{II-ZZ}{2}\right)_{2i-1, 2i+1}}\, ,
\end{align}
\begin{align}
    \overline{\Gamma}_{\leq 2}[\hat{\mathbf{H}}_{3/2}] &= \sum_{2i+1} \left( \frac{II-ZZ}{2} \right)_{ext(2i), ext(2i+2)} \otimes \ketbra{1}{1}_{2i} \otimes \ketbra{1}{1}_{2i+2} \notag \\
    &+ \sum_{2i} \left( \ketbra{1}{1}_{2i-1} \otimes (a_{3/2}Z+b_{3/2}X)_{ext(2i)} \otimes \ketbra{1}{1}_{2i} \otimes \ketbra{1}{1}_{2i+1} \right) \notag\\
    &\textcolor{black}{\equiv \sum_{2i+1} \left( \frac{II-ZZ}{2} \right)_{ext(2i), ext(2i+2)} \notag }\\
    &\textcolor{black}{+ \sum_{2i} \left( \ketbra{1}{1}_{2i-1} \otimes (a_{3/2}Z+b_{3/2}X)_{ext(2i)} \otimes \ketbra{1}{1}_{2i+1} \right) }\, .
\end{align}

\begin{figure}[ht!]
    \centering
    \includegraphics[width=1\linewidth]{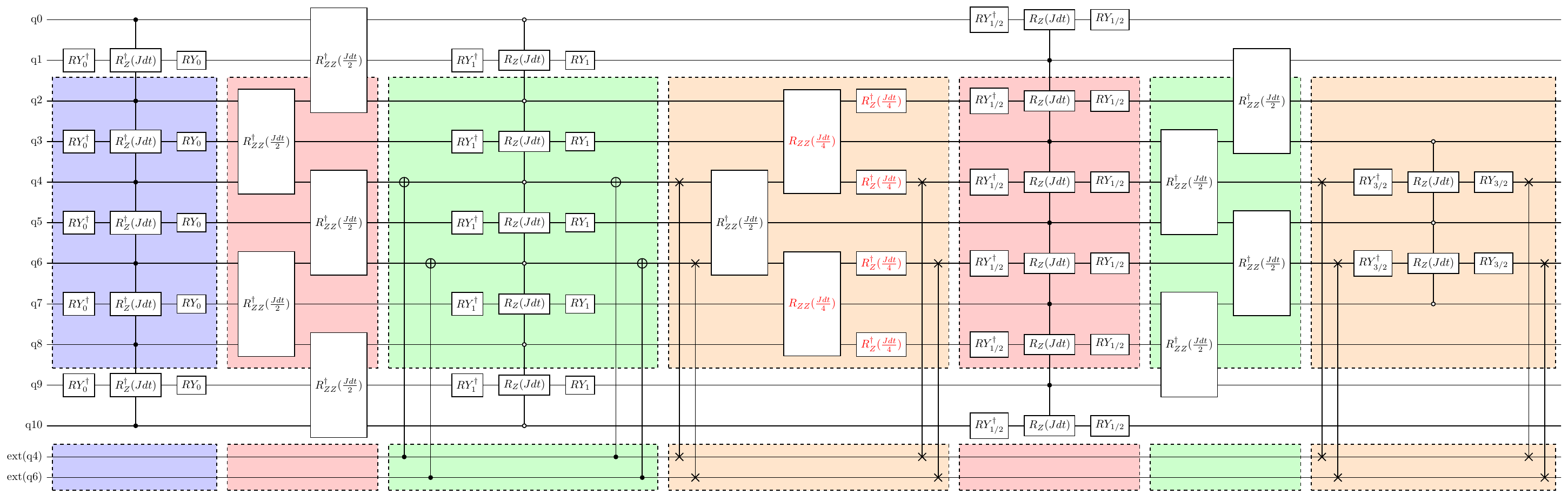}
    \caption{Trotter step implementation in the truncated subspace with $\bar S< 2$. The extra qubit registers required to encode the subspace are indicated by $ext(i)$ and are isolated from the rest of the circuit.}
    \label{fig:tikz_circuit_0020}
\end{figure}

Similar to what was done in \cref{sec:time_evolution_truncated_hamiltonians}, we can write a quantum circuit implementing a single Trotter step within this subspace. After the previous simplifications the resulting circuit is similar to the one of \cref{fig:tikz_circuit_0015_old_new_labels}, with an additional layer accounting for the term $\overline{\Gamma}^{csf}_{\leq 2}[\hat{\mathbf{H}}_{3/2}]$, highlighted in orange in \cref{fig:tikz_circuit_0020}. Note that this circuit requires one additional qubits for all even sites, denoted by the qubit registers $ext(i)$ which are separated from the rest of the circuit to highlight the similarities with previous implementations in smaller subspaces.

\section{Preparing ground state approximations in the triplet subspace}

A well-known analytical result on the spin-$1/2$ antiferromagnetic Heisenberg chain is that the true ground state of the system is in a global singlet configuration~\citep{Bethe1931ZurTheorieMetalle}.
For this reason, we focused in this work mainly on singlet wave functions.
However, states with different spin multiplicity can be naturally targeted using spin-adapted bases.  This would enable, for instance, to address questions such as Haldane's conjecture on the singlet-triplet gap of the infinite spin-$1$ chain~\citep{Haldane1983ContinuumDynamics1D}. 

As already discussed in \cref{sec:spin_adapted_bases}, the commutativity of the total spin operator with the Hamiltonian of the system translates into the conservation of the initial state total spin upon exact time-evolution. A common strategy for preparing approximations of the minimal-energy triplet state is to repeat the non-adapted adiabatic state preparation approach of \cref{sec:adiabatic_ground_state_preparation}, using as initial state an $S=1$ eigenstate expressed in the $\hat{\mathbf{s}}_z$-computational basis. However, the Trotter-induced errors as well as the hardware noise when implementing this strategy can lead to symmetry contamination mixing the different symmetry sectors.

\begin{figure}[ht!]
    \centering
    \includegraphics[width=\linewidth]{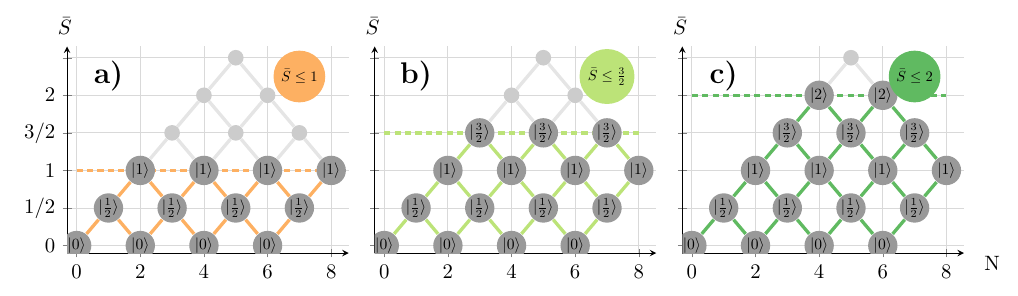}
    \caption{Yamaguchi-Kotani representation of the total-spin eigenstates for a chain of 8 sites with $S=1$. Each total-spin eigenstate is associated with a spin-path connecting the origin to the node with coordinates $(N,S)=(8,1)$. From left to right, truncated symmetry subspaces for the three truncation levels $\bar{S}_{trunc}=\{1, 3/2, 2\}$ in orange, light green and dark green respectively.}
    \label{fig:spin_paths_triplet}
\end{figure}

Our alternative strategy proposed in \cref{sec:adiabatic_ground_state_preparation} can be applied to different symmetry sectors by simply changing the boundary conditions. In particular, we focus on the total spin eigenstates in the triplet configuration, shown for the case $N=8$ in \cref{fig:spin_paths_triplet}, using the Yamaguchi-Kotani representation. The eigenstates of the total spin operator with eigenvalue $S=1$ are mapped to spin-path connecting the origin $(0,0)$ with the final coordinates $(N,S)=(8,1)$. Importantly, the truncation of the symmetry subspaces based on the intermediate total spin can be applied in a similar manner as that of the singlet subspace. The truncated subspaces corresponding the the truncation $\bar{S}_{trunc}\in\{1, 3/2, 2\}$ are also depicted in \cref{fig:spin_paths_triplet} in orange, light green and dark green respectively.

To demonstrate the scalability of our approach also for triplet states, we extend the study of the 16-site Heisenberg chain in \cref{sec:adiabatic_ground_state_preparation}. The encoding of the physical degrees of freedom for the triplet subspace is done using 7 qubits for the $(\bar{S}\leq 1)$-subspace, 14 qubits for the $(\bar{S}\leq 3/2)$-subspace, and 20 qubits for the $(\bar{S}\leq 2)$-subspace. Note that these numbers are larger than the ones obtained in the singlet subspace, requiring 7, 13, and 18 qubits respectively, although the number of physical triplet configurations is reduced. 

\begin{figure}[ht!]
    \centering
    \includegraphics[width=\linewidth]{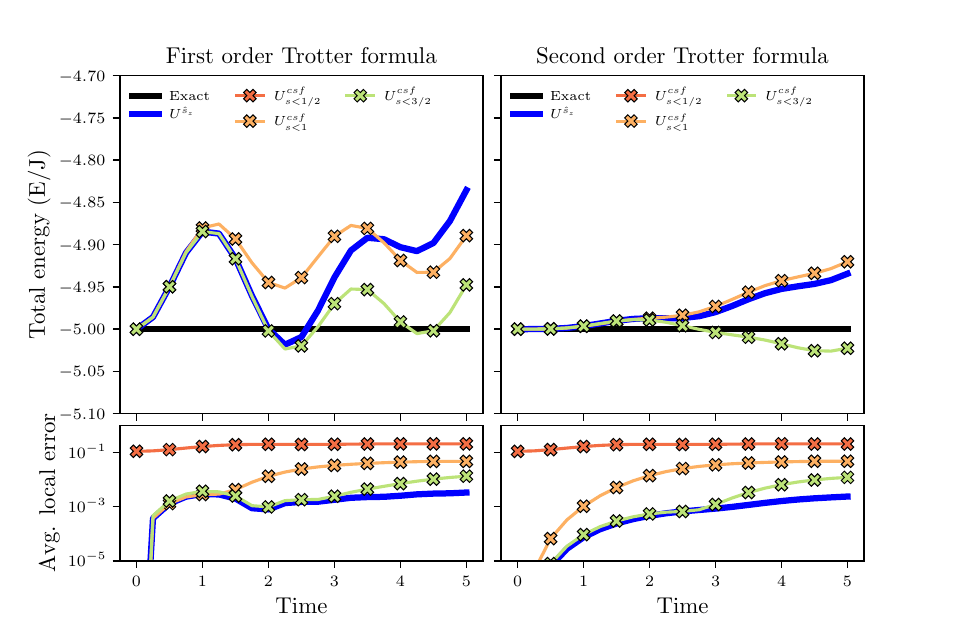}
    \caption{Upper row: Time evolution of the total energy of the initial state $\ket{0,\frac{1}{2},0,\dots,0,\frac{1}{2},1}$ for a 16-site Heisenberg chain in the triplet symetry sector. Lower row: Average absolute error of the time-evolved local energies as a function of time for all unitaries considered in this work. The time-evolved observables are evaluated for the non-spin adapted unitary circuit (blue line) as well as our hierarchy of approximation unitaries in the truncated spin-adapted subspaces (red, orange, and light green crosses), for both the first- and second-order Trotter formulas with $N_L=10$ layer repetitions, as done in \cref{fig:FIG10_combined_plot_total_energy_avg_error} for the singlet subspace.  Note that the largest $(\bar{S}<2)$ subspace for the triplet sector is not included in this figure, because it is not used in this section.}
    \label{fig:FIG10S1_combined_plot_total_energy_avg_error}
\end{figure}

\begin{figure}[ht!]
    \centering
    \includegraphics[width=\linewidth]{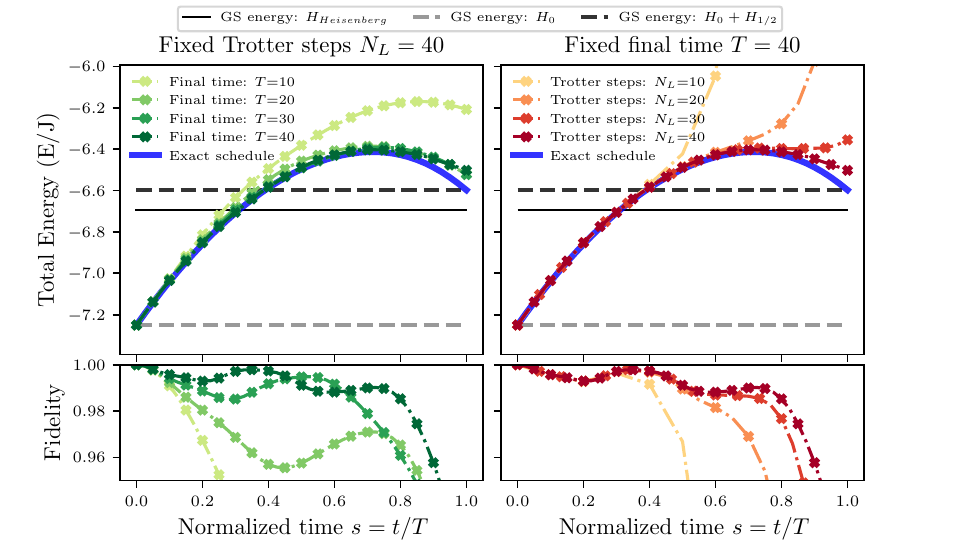}
    \caption{Adiabatic schedule for the 16-site Heisenberg chain in the triplet subspace with truncation $\bar{S}<1$. The total energy and the instantaneous fidelity with respect to the exact schedule are given with the second order trotter formula for: left) a fixed number of Trotter layers $N_L=40$, right) a fixed duration of the adiabatic schedule $T=40$. The horizontal lines represent the exact ground state energies (black line) and the exact ground state energy of the initial (resp. target) Hamiltonian of the adiabatic schedule (dashed gray, resp. black, line).}
    \label{fig:FIG5S1_0010_H0_LH12_duo_energies_fidelities_combined}
\end{figure}

\begin{figure}[ht!]
    \centering
    \includegraphics[width=\linewidth]{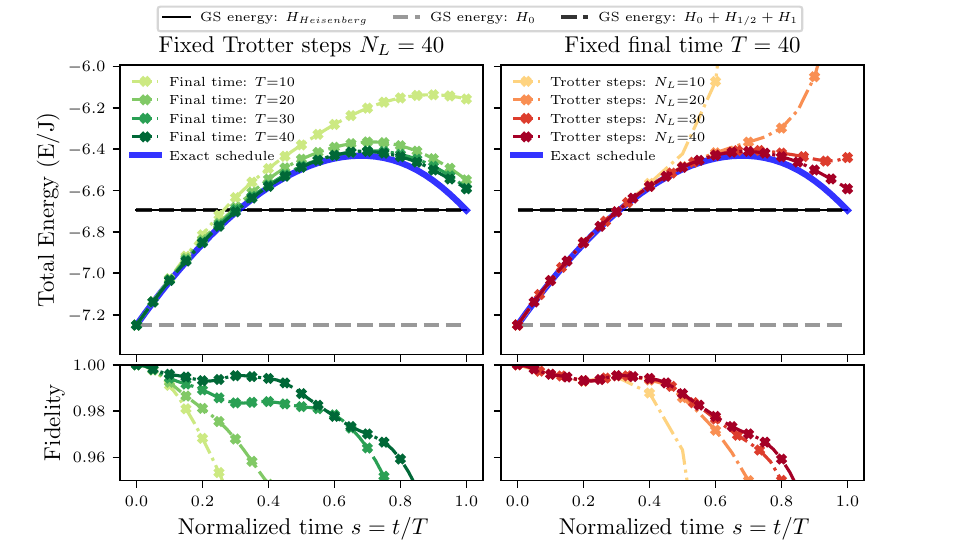}
    \caption{Adiabatic schedule for the 16-site Heisenberg chain in the triplet subspace with truncation $\bar{S}<3/2$. The total energy and the instantaneous fidelity with respect to the exact schedule are given with the second order trotter formula for: left) a fixed number of Trotter layers $N_L=40$, right) a fixed duration of the adiabatic schedule $T=40$. The horizontal lines represent the exact ground state energies (black line) and the exact ground state energy of the initial (resp. target) Hamiltonian of the adiabatic schedule (dashed gray, resp. black, line).}
    \label{fig:FIG5S1_0015_H0_LH12H1_duo_energies_fidelities_combined}
\end{figure}

We use the same adiabatic schedules \cref{eq:Trotter_adiabatic_schedule_0010,eq:Trotter_adiabatic_schedule_0015} defined for the singlet subspace but replace the initial state to be the triplet state denoted as $\ket{u,d,u,d,\ldots,u,u}$ in the step encoding and $\ket{0,\frac{1}{2},0,\dots,0,\frac{1}{2},1}$ in the height encoding. We first show in~\cref{fig:FIG10S1_combined_plot_total_energy_avg_error} that our sequence of circuit unitaries captures well the dynamics of this initial state in the triplet truncated subspaces. We then reproduce the results of the singlet subspace, \cref{fig:0010_H0_LH12_duo_energies_fidelities_combined,fig:0015_H0_LH12H1_duo_energies_fidelities_combined} in the triplet symmetry sector, and report the obtained results in \cref{fig:FIG5S1_0010_H0_LH12_duo_energies_fidelities_combined,fig:FIG5S1_0015_H0_LH12H1_duo_energies_fidelities_combined}.

\end{document}